\documentclass[11pt]{article}
\usepackage[margin=1in]{geometry}
\usepackage{times}

\usepackage{xcolor}
\usepackage[colorlinks=true,citecolor=blue,linkcolor=blue]{hyperref}

\hypersetup{
    colorlinks,
    linkcolor={red!50!black},
    citecolor={blue!50!black},
    urlcolor={blue!80!black}
}

\usepackage[]{amsmath,amssymb,amsfonts,latexsym,amsthm,enumerate,fullpage,xcolor,cite,bbm}
\usepackage[nameinlink, noabbrev, capitalize]{cleveref}
\usepackage{comment}
\usepackage{nicefrac}
\crefname{prop}{Proposition}{Propositions}
\crefname{ineq}{inequality}{inequalities}
\creflabelformat{ineq}{#2(#1)#3}

\newtheorem{theorem}{Theorem}
\newtheorem{lemma}{Lemma}

\newtheorem{fact}{Fact}

\newtheorem{question}{Question}

\newtheorem{corollary}{Corollary}
\newtheorem{definition}{Definition}

\newtheorem{remark}{Remark}

\crefname{THM}{Theorem}{Theorems}

\newtheorem{innercustomthm}{Theorem}
\newenvironment{customthm}[1]
  {\renewcommand\theinnercustomthm{#1}\innercustomthm}
  {\endinnercustomthm}

\usepackage{subcaption}
\usepackage{graphicx}
\usepackage{tikz}
\usepackage{color, colortbl}
\definecolor{LightCyan}{rgb}{0.88,1,1}
\definecolor{Gray}{gray}{0.9}
\usetikzlibrary{positioning}
\usetikzlibrary{decorations.pathmorphing}
\usetikzlibrary{automata,positioning}
\usetikzlibrary{arrows}
\usetikzlibrary{arrows.meta}
\usetikzlibrary{calc}

\newcommand{\E}{\mathbb{E}}
\newcommand{\N}{\mathbb{N}}

\newcommand{\F}{\mathbb{F}}

\newcommand{\supp}{\text{support}}
\newcommand{\pr}{{\prime}}

\newcommand{\U}{\mathbf{U}}
\newcommand{\X}{\mathbf{X}}
\newcommand{\Y}{\mathbf{Y}}

\newcommand{\B}{\mathbf{B}}

\newcommand{\T}{\mathbf{T}}

\newcommand{\bits}{\{0,1\}}
\newcommand{\zo}{\{0,1\}}

\newcommand{\Ext}{\mathsf{Ext}}
\newcommand{\Disp}{\mathsf{Disp}}

\newcommand{\Q}{\mathbf{Q}}

\newcommand{\eps}{\varepsilon}
\newcommand{\AC}{\mathsf{AC}}

\newcommand{\cC}{\mathcal{C}}

\newcommand{\minH}{H_\infty}

\DeclareMathOperator*{\argmax}{\arg\!\max}
\DeclareMathOperator*{\bias}{bias}
\DeclareMathOperator*{\corr}{corr}

\usepackage{subfiles}
\newcommand{\dobib}{
    \bibliographystyle{alpha}
    \bibliography{references} 
}

\begin{document}
\renewcommand{\dobib}{}

\title{Low-Degree Polynomials Extract from Local Sources}
\author{Omar Alrabiah\\ UC Berkeley\\ \texttt{oalrabiah@berkeley.edu}\and Eshan Chattopadhyay\thanks{Supported by NSF CAREER Award 2045576.}\\ Cornell University\\ \texttt{eshan@cs.cornell.edu} \and Jesse Goodman\footnotemark[1]\\ Cornell University\\ \texttt{jpmgoodman@cs.cornell.edu} \and Xin Li\thanks{Supported by NSF CAREER Award CCF-1845349.}\\ Johns Hopkins University\\ \texttt{lixints@cs.jhu.edu} \and Jo\~ao Ribeiro\thanks{Research supported in part by the NSF grants CCF-1814603 and CCF-2107347 and by the following grants of Vipul Goyal: the NSF award 1916939, DARPA SIEVE program, a gift from
Ripple, a DoE NETL award, a JP Morgan Faculty Fellowship, a PNC center for financial services
innovation award, and a Cylab seed funding award.}\\ Carnegie Mellon University\\ \texttt{jlourenc@cs.cmu.edu}}
\date{}

\begin{titlepage}
\maketitle
\begin{abstract}
  
We continue a line of work on extracting random bits from weak sources that are generated by simple processes. We focus on the model of locally samplable sources, where each bit in the source depends on a small number of (hidden) uniformly random input bits. Also known as local sources, this model was introduced by De and Watson (TOCT 2012) and Viola (SICOMP 2014), and is closely related to sources generated by \(\AC^0\) circuits and bounded-width branching programs. In particular, extractors for local sources also work for sources generated by these classical computational models.

Despite being introduced a decade ago, little progress has been made on improving the entropy requirement for extracting from local sources. The current best explicit extractors require entropy \(n^{1/2}\), and follow via a reduction to affine extractors. To start, we prove a barrier showing that one cannot hope to improve this entropy requirement via a black-box reduction of this form. In particular, new techniques are needed.

In our main result, we seek to answer whether low-degree polynomials (over \(\F_2\)) hold potential for breaking this barrier. We answer this question in the positive, and fully characterize the power of low-degree polynomials as extractors for local sources. 
More precisely, we show that a random degree \(r\) polynomial is a low-error extractor for \(n\)-bit local sources with min-entropy \(\Omega(r(n\log n)^{1/r})\), and we show that this is tight.

Our result leverages several new ingredients, which may be of independent interest. Our existential result relies on a new reduction from local sources to a more structured family, known as local non-oblivious bit-fixing sources. To show its tightness, we prove a ``local version'' of a structural result by Cohen and Tal (RANDOM 2015), which relies on a new ``low-weight'' Chevalley-Warning theorem.

\end{abstract}
\thispagestyle{empty}
\end{titlepage}
\pagenumbering{arabic}
\newpage

\section{Introduction}\label{sec:intro}
Randomness is a fundamental resource in many areas of computer science, such as algorithm design and cryptography. However, such tasks often assume access to a source of independent and uniform bits, while real-world physical processes (e.g., electromagnetic noise, timings of user keystrokes) generate randomness that is far from perfect.
This state of affairs motivates the problem of \emph{randomness extraction}. The goal is to design a deterministic function, called an \emph{extractor}, that can distill a (nearly) uniform bit from any source belonging to a certain family.\footnote{In general, one would actually like to distill (from the source) as many uniform bits as possible. However, we focus here on the simpler goal of obtaining just a single uniform bit, as there remain significant barriers even in this most basic setting.}

\begin{definition}[Extractor]\label{def:intro:extractor}
A function $\Ext:\bits^n\to\bits$ is an \emph{extractor} for a family of distributions $\mathcal{X}$ over \(\zo^n\) with error \(\eps\) if, for every \(\X\in\mathcal{X}\),
\begin{equation*}
    \left|\Pr[\Ext(\X)=1]-\frac{1}{2}\right|\leq \eps.
\end{equation*}
We also call \(\Ext\) an \emph{\(\eps\)-extractor for \(\mathcal{X}\).}
\end{definition}

Besides their practical motivation, randomness extractors (and other related pseudorandom objects such as dispersers, condensers, and expander graphs) have deep connections to coding theory, combinatorics, and complexity theory.

In order to construct an extractor for a family \(\mathcal{X}\) of sources, the most general assumption one can make about \(\mathcal{X}\) is that each \(\X\in\mathcal{X}\) has some ``randomness.'' Here, it is typical to measure the randomness of a source \(\X\) by its min-entropy \(H_\infty(\X):=-\log\max_x\Pr[\X=x]\). However, even if we assume each \(\X\in\mathcal{X}\) has a very high amount of this very strong notion of entropy, extraction is still impossible: indeed, one cannot hope to extract from \(\mathcal{X}\) even if each source \(\X\in\mathcal{X}\) is guaranteed to have min-entropy \(k\geq n-1\)~\cite{CG88}. To enable extraction, one must make additional assumptions on the structure of each \(\X\in\mathcal{X}\).

\paragraph*{Extractors for local sources, \(\mathsf{AC^0}\) sources, and small-space sources.}
In a seminal work, Trevisan and Vadhan~\cite{TV00} initiated the study of randomness extraction from sources that can be sampled by ``simple'' processes. In addition to the generality of such sources, it can be argued that they serve a reasonable model of randomness that might actually be found in nature.

More formally, Trevisan and Vadhan studied sources that can be sampled by polynomial size circuits that are given uniform bits as input. However, extracting randomness from this class of sources requires strong computational hardness assumptions.
This motivated De and Watson~\cite{DW12} and Viola~\cite{Vio14} to consider \emph{unconditional} extraction from sources sampled by more restricted, but still natural, circuit families.
To this end, they introduced the notion of \emph{local sources}.
Intuitively, a local source \(\X\) is one that can be sampled by a low-depth circuit with bounded fan-in (a low-complexity process).

\begin{definition}[Local source~\cite{DW12,Vio14}]\label{def:localsources}
A distribution $\X\sim\bits^n$ is a \emph{$d$-local source} if \(\X=g(\U_m)\), where \(\U_m\) is the uniform distribution over \(m\) bits (for some \(m\)), and \(g:\zo^m\to\zo^n\) is a function where each output bit depends on at most \(d\) input bits.
\end{definition}

Local sources are closely connected to other models of sources sampled by simple processes. Viola~\cite{Vio14} proved that every source generated by $\AC^0$ circuits is (close to) a convex combination of local sources with small locality and slightly lower min-entropy. More recently, Chattopadhyay and Goodman~\cite{CG21} showed a similar result for sources generated by bounded-width branching programs~\cite{KRVZ11}. Thus, extractors for local sources also work for sources generated by these classical computational models. In fact, the current state-of-the-art extractors for sources generated by \(\AC^0\) circuits and bounded-width branching programs are extractors for \(1\)-local sources.

\paragraph*{A barrier at \(\sqrt{n}\) min-entropy.}

Despite the applications above and being introduced over a decade ago, little progress has been made on constructing extractors for local sources \cite{DW12,Vio14,Li16}. In particular, all known constructions require min-entropy at least \(\sqrt{n}\), and follow via a reduction to extractors for affine sources (i.e., sources that are uniform over affine subspaces of \(\F_2^n\)). Thus, there appears to be a ``barrier'' at \(\sqrt{n}\) min-entropy, at least when using affine extractors \cite{Vio20}. It is natural to ask how we might break this \(\sqrt{n}\) barrier, which raises the question:
\begin{center}
    \emph{Can affine extractors be used to extract from local sources with min-entropy \(k\ll \sqrt{n}\)?}
\end{center}

As motivation, we start by providing strong evidence that the answer to the above question is negative. In particular, we prove the following.

\begin{customthm}{0}[Barrier result]\label{claim:barrierintro}
It is not possible to extract randomness from $2$-local sources with min-entropy $k\geq\sqrt{n}$ by applying an affine extractor in a black-box manner.
\end{customthm}

Thus, if we would like to construct extractors for local sources with min-entropy significantly below \(\sqrt{n}\), new techniques are needed.

\paragraph*{Towards breaking the \(\sqrt{n}\) barrier using low-degree polynomials.}

While explicit extractors that break the \(\sqrt{n}\) min-entropy barrier for local sources are the end goal, these still seem beyond reach. 
We believe that the next best thing are non-explicit extractors that are of “low complexity.”
Our hope is that such extractors may help us eventually construct truly explicit extractors, as non-explicit extractors are more likely to be easier to derandomize if they belong to a low complexity class. At the same time, such non-explicit extractors may have applications in complexity theory (i.e., since the current state-of-the-art circuit lower bounds are against extractors \cite{li20213}). There is a long line of work~\cite{Vio05,GVW15,CL18,Lu04,Vad04,BogG13,CL18,DY21,HIV21,CT15} on the power of low-complexity computational models for extracting from various families of sources.

We choose to study the class of \emph{low-degree \(\F_2\)-polynomials} as our low-complexity computational model. In particular, we ask whether (non-explicit) low-degree polynomials can help break the \(\sqrt{n}\) barrier for extracting from local sources, and more generally we seek to answer the following question:

\begin{question}\label{q:mainq}
    How powerful are low-degree $\F_2$-polynomials as extractors for local sources?
\end{question}

Beyond being a natural algebraic class, low-degree $\F_2$-polynomials have a natural combinatorial interpretation.
We can represent a degree-$2$ $\F_2$-polynomial $f$ as a graph $G_f$ on $n$ vertices, with edges representing monomials included in $f$.
Then, $f$ being a good extractor for local sources translates into a parity constraint on the number of edges in certain induced subgraphs of $G_f$.
Likewise, a degree-$3$ $\F_2$-polynomial can be represented as a 3-hypergraph, and so on. Given the correspondence between low-degree polynomials and hypergraphs with small edge sizes, we hope that tools from combinatorics can be leveraged to make our constructions explicit and break the $\sqrt{n}$ min-entropy barrier for extracting from local sources (which would also give improved extractors for small-space sources).

Our motivation to study low-degree polynomials as our ``low complexity'' model also comes from the work of Cohen and Tal \cite{CT15}, which studied the same question in the context of \emph{affine} sources. In their work, they showed that there exist degree-$r$ $\F_2$-polynomials that extract from affine sources with min-entropy $O(r n^{\frac{1}{r-1}})$, and that this is tight. In order to answer \cref{q:mainq}, we aim to provide a local source analogue of this result.

\subsection{Summary of our results}\label{sec:summary}

In this paper, we fully characterize the power of low-degree polynomials as extractors for local sources, answering \cref{q:mainq} and proving a local-source analogue of Cohen-Tal. Along the way, we rely on several new ingredients which may be of independent interest. We present these results in \cref{subsub:first-part-summary} and \cref{subsub:second-part-summary}, respectively.

\subsubsection{Main result}\label{subsub:first-part-summary}

Our main result gives a tight characterization of the power of low-degree polynomials as extractors for local sources. We state it formally, below.

\begin{customthm}{1}[Main result]\label{thm:absolute-main}
For every \(d\in\N\) and \(r\geq2\), there are constants \(C,c>0\) such that the following holds. For every \(n\in\N\) there exists a (not necessarily explicit) degree \(\leq r\) polynomial \(f\in\F_2[x_1,\dots,x_n]\) that is a \((2^{-ck})\)-extractor for \(d\)-local sources with min-entropy
\[
k\geq C (n\log n)^{1/r},
\]
but for every degree \(\leq r\) polynomial \(g\in\F_2[x_1,\dots,x_n]\) there exists a \(d\)-local source with min-entropy
\[
k\geq c (n\log n)^{1/r}
\]
on which it is constant.
\end{customthm}

\cref{thm:absolute-main} implies that degree-\(3\) polynomials are already enough to extract from min-entropy \(k=O((n\log n)^{1/3})\), which (non-explicitly) breaks the \(\sqrt{n}\) min-entropy barrier of previous techniques (\cref{claim:barrierintro}). Furthermore, given known reductions from \(\AC^0\) sources and small-space sources to local sources~\cite{Vio14,CG21}, it also follows that low-degree polynomials can be used to break the current min-entropy barriers for extracting from these other models of weak sources. Finally, \cref{thm:absolute-main} can be viewed as a generalization of a result by Viola \cite{Vio16}, who showed the existence of quadratic functions \(f\) such that \(\AC^0\) cannot sample \((\U_n,f(\U_n))\): our result shows that there exist degree \(\leq r\) polynomials such that \(\AC^0\) cannot sample \((\X,f(\X))\), where \(\X\) can be \emph{any} distribution with min-entropy \(k\geq Cn^{1/r+\delta}\) for a large enough \(C\) and arbitrarily small \(\delta>0\).

While \cref{thm:absolute-main} is stated for constant locality \(d\) and constant degree \(r\), we actually prove stronger results that hold for superconstant \(d,r\). In particular, \cref{thm:absolute-main} follows immediately from the following two results, which provide upper and lower bounds on the entropy required to extract from \(d\)-local sources using a degree \(\leq r\) polynomials (where \(d,r\) need not be constant).

\begin{customthm}{1.1}[Technical version of \cref{thm:absolute-main}, Upper Bound]\label{thm:localextintro}
There are universal constants \(C,c>0\) such that for all \(n,d,r\in\N\), the following holds. With probability at least \(0.99\) over the choice of a random degree \(\leq r\) polynomial \(f\in\F_2[x_1,\dots,x_n]\), it holds that \(f\) is an \(\eps\)-extractor for \(d\)-local sources with min-entropy \(k\), for any
\[
k\geq C2^dd^2r\cdot(2^dn\log n)^{1/r},
\]
where \(\eps=2^{-\frac{ck}{r^32^dd^2}}\). 
\end{customthm}

\begin{remark}[Informal version of \cref{thm:formal-version-of-main-remark}]\label{rem:main-remark-here}
If instead of extractors we aim to construct \emph{dispersers},\footnote{A function $\Disp:\bits^n\to\bits$ is a \emph{disperser for a class of sources $\cC$} if the support of $\Disp(\X)$ is $\bits$ for all sources $\X\in\cC$.} then we are able to improve \cref{thm:localextintro} to hold for min-entropy $k\geq Cd^3r\cdot(n\log n)^{1/r}$.
\end{remark}

\begin{customthm}{1.2}[Technical version of \cref{thm:absolute-main}, Lower Bound]\label{thm:localtightintro}
There exists a universal constant \(c>0\) such that for all \(n,d,r\in\N\) with \(2\leq r\leq c\log(n)\) and \(d\leq n^{\frac{1}{r-1}-2^{-10r}}/\log(n)\), the following holds. For any degree \(\leq r\) polynomial \(f\in\F_2[x_1,\dots,x_n]\), there is a \(d\)-local source \(\X\sim\zo^n\) with min-entropy at least
\[
k\geq cr (dn\log n)^{1/r}
\]
such that $f(\X)$ is constant.
\end{customthm}

\subsubsection{Key new ingredients}\label{subsub:second-part-summary}

Our main result follows from a collection of new ingredients, which may be of independent interest. In order to prove our upper bound on min-entropy (\cref{thm:localextintro}), we prove a new reduction from \(d\)-local sources to \emph{$d$-local non-oblivious bit fixing (NOBF) sources}. Informally, a $d$-local NOBF source $\mathbf{X} \sim \{0,1\}^n$ of min-entropy $k'$ is a source that has $k'$ uniform independent bits, with all other bits depending on at most $d$ of the $k'$ bits.

\begin{customthm}{2}[Reduction from \(d\)-local sources to \(d\)-local NOBF sources]
\label{thm:red-to-nobf}
There exists a universal constant \(c>0\) such that for any \(n,k,d\in\N\), the following holds. Let \(\X\sim\zo^n\) be a \(d\)-local source with min-entropy \(\geq k\). Then \(\X\) is \(\eps\)-close to a convex combination of \(d\)-local NOBF sources with min-entropy \(\geq k^\pr\), where \(\eps=2^{-ck^\pr}\) and
\[
k^\pr=\frac{ck}{2^d d^2}.
\]
\end{customthm}

The family of \(d\)-local NOBF sources, introduced in~\cite{CGGL20}, is a significant specialization of \(d\)-local sources. The above reduction shows that, at least for constant locality \(d\), we can just focus on extracting from this simpler class - even in future explicit constructions.

To prove our lower bound on min-entropy (\cref{thm:localtightintro}), we actually prove this lower bound for the special class of \(d\)-local sources known as \emph{$d$-local affine sources}: such a source \(\X\sim\F_2^n\) is uniform over a \(d\)-local affine subspace \(X\subseteq\F_2^n\), which is a special type of affine subspace that admits a basis \(v_1,\dots,v_k\in\F_2^n\) where each coordinate \(i\in[n]\) holds the value \(1\) in at most \(d\) of these vectors. For this special class of sources, our lower bound is actually tight, and can be viewed as a ``local'' version of a result by Cohen and Tal~\cite{CT15}.

\begin{customthm}{3}[Local version of Cohen-Tal]\label{thm:local-cohen-tal}
There exist universal constants \(C,c>0\) such that for every \(n,r,d\in\N\) such that \(2\leq r\leq c\log(n)\) and \(d\leq n^{\frac{1}{r-1}-2^{-10r}}/\log(n)\), the following holds. For any degree \(\leq r\) polynomial \(f\in\F_2[x_1,\dots,x_n]\), there exists a \(d\)-local affine subspace \(X\subseteq\F_2^n\) of dimension
\[
k\geq cr (d n\log n)^{1/r}
\]
on which \(f\) is constant.

This is tight: there exists a degree \(\leq r\) polynomial \(g\in\F_2[x_1,\dots,x_n]\) which is an extractor for \(d\)-local affine sources of dimension \(k\geq Cr(dn\log n)^{1/r}\), which has error \(\eps=2^{-ck/r}\).
\end{customthm}

We prove \cref{thm:local-cohen-tal} by extending the techniques of Cohen and Tal \cite{CT15}, while leveraging a key new ingredient: a ``low-weight'' Chevalley-Warning theorem. This result, which may be of independent interest, shows that any small system of low-degree polynomials admits a (nontrivial) solution of low Hamming weight.

\begin{customthm}{4}[Low-weight Chevalley-Warning]
\label{thm:low-wt-cw}
Let \(\{f_i\}\subseteq\F_2[x_1,\dots,x_n]\) be a set of polynomials with linear degree\footnote{The \emph{linear degree} $D$ is the sum of the degrees of the $f_i$'s which have degree \(1\).} at most \(D\) and nonlinear degree\footnote{The \emph{nonlinear degree} is the sum of the degrees of the $f_i$'s which have degree at least $2$.} at most \(\Delta\), such that \(0\) is a common solution and \(D+\Delta<n\). Then there is a common solution \(x\neq 0\) with Hamming weight
\[
    w\leq 8\Delta + 8D/\log(n/D)+8.
\]
\end{customthm}

\subsection{Open problems}

Our work leaves open several interesting avenues for future work. We highlight three of them here:
\begin{itemize}
    \item For any constant $r \ge 2$, construct an explicit $\F_2$-polynomial of degree $\leq r$ that extracts from 2-local sources of min-entropy $o(n)$.
    
    \item In \cref{thm:red-to-nobf}, we showed a reduction from \(d\)-local sources of min-entropy $k$ to \(d\)-local non-oblivious bit fixing (NOBF) sources of min-entropy \(\Omega(k/(2^dd^2))\). It would be interesting to show a reduction from a \(d\)-local source of min-entropy \(k\) to a \(d\)-local NOBF source of min-entropy \(\Omega(k/\text{poly}(d))\) (or show that such a reduction is impossible).
    
    \item \cref{thm:low-wt-cw} shows that if a collection of low-degree \(n\)-variate polynomials of linear degree $D$ and nonlinear degree $\Delta$ has the zero vector as a solution (and \(D+\Delta<n\)), then there exists a nonzero solution of weight at most $O(\Delta + D/\log(n/D))$. When $\Delta = 0$, this becomes asymptotically tight by the Hamming bound. Moreover, when $D = 0$, this will also be tight by picking the polynomial $f(x) = \sum_{1 \le |S| \le \Delta}{\prod_{i \in S}{x_i}}$. However, if we had $\Delta/2$ quadratic polynomials, will the upper bound of $O(\Delta)$ still be tight?
\end{itemize}

\section{Overview of our techniques}\label{sec:overview}

In this section, we provide an overview of the techniques that go into our three main results:

\begin{itemize}
    \item An entropy \emph{upper bound} for low-degree extraction from local sources (\cref{thm:localextintro}).
    \item An entropy \emph{lower bound} for low-degree extraction from local sources (\cref{thm:localtightintro}).
    \item A \emph{barrier} for extracting from local sources using black-box affine extractors (\cref{claim:barrierintro}).
\end{itemize}

Along the way, we will overview the several new key ingredients (Theorems~\ref{thm:red-to-nobf},~\ref{thm:local-cohen-tal}, and~\ref{thm:low-wt-cw}) that go into these main results.

\subsection{Upper bounds}\label{sec:techoverUB}

We begin by discussing our entropy upper bounds for low-degree extraction from local sources. By this, we mean that we upper bound the entropy \emph{requirement} for extracting from \(d\)-local sources using degree \(\leq r\) polynomials. In other words, we show that low-degree polynomials extract from local sources.

\subsubsection{Low-degree extractors for local sources}
We start by sketching the techniques behind our main upper bound (\cref{thm:localextintro}), which shows that most degree \(\leq r\) polynomials are low-error extractors for \(d\)-local sources with min-entropy at least \[k=O(2^dd^2r\cdot(2^dn\log n)^{1/r}).\]

\paragraph*{A strawman application of the probabilistic method.}
A natural first attempt at proving our result would use a standard application of the probabilistic method, which looks something like the following. First, let \(f\in\F_2[x_1,\dots,x_n]\) be a uniformly random polynomial of degree \(\leq r\), meaning that each monomial of size \(\leq r\) is included in \(f\) with probability \(1/2\). Then, we let \(\mathcal{X}\) be the family of \(d\)-local sources over \(\zo^n\), each with min-entropy at least \(k\). To prove that most of these polynomials are low-error extractors for this family, a standard application of the probabilistic method would suggest that we:
\begin{enumerate}
    \item Prove that \(f\) is an extractor for a single \(\X\in\mathcal{X}\) with extremely high probability.
    \item Show that the family \(\mathcal{X}\) does not contain too many sources.
    \item Conclude, via the union bound, that \(f\) is an extractor for \emph{every} \(\X\in\mathcal{X}\) with high probability.
\end{enumerate}

It is not too hard to complete Steps \(2\) and \(3\) in the above framework, but Step \(1\) turns out to be much more challenging. To see why, let us consider an arbitrary \(d\)-local source \(\X\sim\zo^n\) with min-entropy at least \(k\). By definition of \(d\)-local source, there exists some \(m\in\N\) and functions \(g_1,\dots,g_n:\zo^m\to\zo\) such that each \(g_i\) depends on just \(d\) of its inputs, and such that given a uniform \(\Y\sim\zo^m\), we have
\[
\X=(g_1(\Y),g_2(\Y),\dots,g_n(\Y)).
\]
Now, we want to argue that a random degree \(\leq r\) polynomial \(f\in\F_2[x_1,\dots,x_n]\) is a low-error extractor for \(\X\). To do so, consider the function \(F:\zo^m\to\zo\) defined as
\[
F(y_1,\dots,y_m):=(f\circ g)(y_1,\dots,y_m)=f\left(g_1(y_1,\dots,y_m),\dots,g_n(y_1,\dots,y_m)\right).
\]
Notice that by the definition of \(\X\) and by \cref{def:intro:extractor} of extractor, we know that \(f\) is an extractor for \(\X\) with error \(\eps\) if
\[
|\bias(F)|:=\left|\Pr_{y\sim\U_m}[F(y)=1]-\Pr_{y\sim\U_m}[F(y)=0]\right|\leq2\eps.
\]
Thus, to argue that a random degree \(\leq r\) polynomial \(f\in \F_2[x_1,\dots,x_n]\) is a low-error extractor for \(\X\), it suffices to argue that the function \(F=f\circ g\) has low bias (with high probability over the selection of \(f\)).

Of course, the question now becomes: how can we ensure that \(F\) has low bias? We can start by noticing some properties of \(F\). First, we know \(F=f\circ g\), where \(f\) is a random degree \(\leq r\) polynomial and \(g\) is a fixed function where each output bit depends on \(\leq d\) input bits. Thus, it is not hard to argue that \(F\) will have degree \(\leq rd\). Furthermore, since \(f\) is random and \(g\) is fixed, one may hope to argue that \(F\) is a \emph{uniformly random} polynomial of degree \(\leq rd\): in this case we would be done, since it is well-known that uniformly random low-degree polynomials have extremely low bias (with extremely high probability) \cite{low-bias-polys}.

Unfortunately, it is too much to hope that \(F\) is a uniformly random low-degree polynomial. Indeed, it is not hard to see that the distribution of \(F\) over degree \(\leq rd\) polynomials depends heavily on the exact selection of \(g\). Furthermore, for most selections of \(g\), the random function \(F\) is \emph{not} uniformly distributed over degree \(\leq t\) polynomials for \emph{any} \(t\).

Thus, there is no obvious way to apply \cite{low-bias-polys} in order to argue that \(F\) will have low bias. To proceed, it seems like we will somehow need to argue that the distribution of \(F\) over low-degree polynomials is guaranteed to have some specific \emph{structure}, and then somehow argue that a random polynomial from any such structured distribution is guaranteed to have low-bias. Each of these steps seems quite challenging.

\paragraph*{Reductions to the rescue.}
As it turns out, there is a simple trick we can use to greatly simplify the above approach. The key idea is to \emph{reduce} local sources to a simpler class of sources. Given two familes \(\mathcal{X},\mathcal{Y}\) of distributions over \(\zo^n\), we say that \(\mathcal{X}\) \emph{reduces} to \(\mathcal{Y}\) if each \(\X\in\mathcal{X}\) is (close to) a convex combination of \(\Y\in\mathcal{Y}\).\footnote{By this we mean that each \(\X\in\mathcal{X}\) can be written in the form \(\X=\sum_ip_i\Y_i\), where each \(\Y_i\in\mathcal{Y}\), \(\sum_ip_i=1\), and \(\X\) samples from \(\Y_i\) with probability \(p_i\).} Reductions are extremely useful, because of the following well-known fact: if \(\mathcal{X}\) reduces to \(\mathcal{Y}\), and \(f:\zo^n\to\zo\) is an extractor for \(\mathcal{Y}\), then \emph{\(f\) is also an extractor for \(\mathcal{X}\)}.

Thus, in order to show that low-degree polynomials extract from \(d\)-local sources, a key new ingredient we use is a reduction from \(d\)-local sources to a simpler class of sources called \emph{\(d\)-local non-oblivious bit-fixing (NOBF) sources} \cite{CGGL20}. Using the above discussion, it then suffices to show that low-degree polynomials extract from \(d\)-local NOBF sources. Thus, we proceed by:
\begin{enumerate}
    \item Defining local NOBF sources, and showing how we can appropriately tailor our previous attempt at the probabilistic method so that it works for local NOBF sources.
    \item Providing a new reduction from local sources to local NOBF sources.
\end{enumerate}

\paragraph*{Low-degree extractors for local NOBF sources.}
A \emph{\(d\)-local NOBF source} \(\X\sim\zo^n\) is a natural specialization of a \(d\)-local source where the entropic bits of the source must show up ``in plain sight'' somewhere in the source.\footnote{The relationship between local sources and local \emph{NOBF} sources is not dissimilar to the relationship between error-correcting codes and \emph{systematic} error-correcting codes.} More formally, a \(d\)-local NOBF source with min-entropy \(k\) is a random variable \(\X\sim\zo^n\) for which there exist functions \(g_1,\dots,g_n:\zo^k\to\zo\) such that the following holds: each \(g_i,i\in[n]\) depends on \(\leq d\) input bits; for every \(i\in[k]\) there is some \(i^\pr\in[n]\) such that \(g_{i^\pr}(y)=y_i\); and for uniform \(\Y\sim\zo^k\) we have
\[
\X=(g_1(\Y),g_2(\Y),\dots,g_n(\Y)).
\]
In other words, some \(k\) ``good'' bits in \(\X\) are uniform, and the remaining \(n-k\) ``bad'' bits are \(d\)-local functions of the good bits.

We must now show that a random degree \(\leq r\) polynomial \(f\in\F_2[x_1,\dots,x_n]\) extracts from \(d\)-local NOBF sources with entropy \(k\). As in our strawman application of the probabilistic method, consider an arbitary \(d\)-local NOBF source \(\X=(g_1(\Y),\dots,g_n(\Y))\) and let \(f\in\F_2[x_1,\dots,x_n]\) be a uniformly random degree \(\leq r\) polynomial. To show that \(f\) extracts from all \(d\)-local NOBF sources, recall that we just need to show that the function \(F:\zo^k\to\zo\) defined as
\[
F(y):=(f\circ g)(y)=f\left(g_1(y),\dots,g_n(y)\right)
\]
has extremely low bias with extremely high probability. Furthermore, recall that if we can show that \(F\) itself is a uniformly random low-degree polynomial, then we know via \cite{low-bias-polys} that this is true.

It is still too much to hope that \(F\) is a uniform low-degree polynomial, but \(F\) is now ``close enough in structure'' to one so that we can make this work. To see why, we can first assume without loss of generality (by definition of local NOBF source) that \(g_1(y)=y_1,\dots,g_k(y)=y_k\). Thus, we can define \(c_S\sim\zo\) as an independent uniform bit (for each \(S\subseteq[n]\) of size \(\leq r\)) and write
\begin{align*}
F(y) = \sum_{S\subseteq[k]:|S|\leq r}c_S\prod_{i\in S}g_i(y) + \sum_{S\subseteq[n]:|S|\leq r, S\not\subseteq[k]}c_S\prod_{i\in S}g_i(y) = A(y) + B(y),
\end{align*}
where \(A\in\F_2[y_1,\dots,y_k]\) is a uniformly random polynomial of degree \(\leq r\), and \(B\in\F_2[y_1,\dots,y_k]\) is a random polynomial whose selection of monomials is \emph{not uniformly random}, but is nevertheless \emph{independent} of the selections made by \(A\).

Thus, to show that \(F\) has extremely low bias, it suffices to show that \(A+B\) has extremely low bias. And to show that \(A+B\) has extremely low bias, it suffices to show that \(A+B^\pr\) has low bias for any fixed polynomial \(B^\pr\) induced by fixing the random monomials selected by \(B\).

To conclude, we actually show something stronger: recalling that \(A\in\F_2[y_1,\dots,y_k]\) is a uniformly random polynomial of degree \(\leq r\), we show that: for \emph{any} fixed function \(B^\ast:\zo^k\to\zo\), it holds that \(A+B^\ast\) has low bias. This does not follow immediately from the result \cite{low-bias-polys} that a random low-degree polynomial has low bias: indeed, it is more general, since \cite{low-bias-polys} is the special case where \(B^\ast=0\). However, it \emph{does} follow immediately from known upper bounds on the list size of Reed-Muller codes \cite{KLP12}.

In the language we are using here, an upper bound on the list size of a Reed-Muller code is equivalent to saying that for any fixed function \(D\in\F_2[x_1,\dots,x_k]\), \(A\) will differ from \(D\) on many inputs, with very high probability. Thus, such bounds tell us that \(A\) differs from \(B^\ast\) on many inputs with very high probability, and \(A\) differs from \(1+B^\ast\) (or rather, \emph{equals} \(B^\ast\)) on many inputs with very high probability. In other words, \(A\) is completely uncorrelated with \(B^\ast\), meaning that \(\bias(A+B^\ast)\) is extremely small, as desired.

Thus a random low-degree polynomial \(f\) extracts from the \(d\)-local NOBF source \(\X\) with min-entropy \(k\) with very high probability. In other words, it fails to do so with some very small probability \(\delta=\delta(k)\) which decreases rapidly as \(k\) grows. By applying the union bound, we get that \(f\) extracts from the entire family \(\mathcal{X}\) of \(d\)-local NOBF sources, provided \(\delta(k)\cdot|\mathcal{X}|\ll 1\). All that remains is to upper bound the size of \(\mathcal{X}\), which can easily be done using the \(d\)-locality of the sources.

\paragraph*{A reduction to local NOBF sources.}
Above, we saw that random low-degree polynomials extract from \emph{local NOBF sources}. To complete the proof that they also extract from more general \emph{local sources}, recall that we need to provide a reduction from local sources to local NOBF sources. In other words, we need to show that every \(d\)-local source with min-entropy \(k\) is (close to) a convex combination of \(d\)-local NOBF sources with min-entropy \(k^\pr\approx k\). This is the main key ingredient in our result that low-degree polynomials extract from local sources (\cref{thm:localextintro}).

Our reduction works as follows. First, pick an arbitrary \(d\)-local source \(\X\sim\zo^n\) with min-entropy \(k\). Let \(k^\pr\) be a parameter which is slightly smaller than \(k\), which will be picked later. We start by arguing that \(\X\) is (close to) a convex combination of \(d\)-local NOBF sources where there are \(k^\pr\) good bits, but the good bits may be biased (but not constant).

Towards this end, recall that \(\X=(g_1(\Y),\dots,g_n(\Y))\) for some \(d\)-local functions \(g_1,\dots,g_n:\zo^m\to\zo\) and uniform \(\Y\sim\zo^m\). The key idea is to consider the \emph{largest possible set} \(T\subseteq[n]\) of ``good bits,'' i.e., such that \(\{\X_i\}_{i\in T}\) are independent (and none are constants). Then, we let \(T^\pr\subseteq[m]\) be the bits of \(\Y\) on which \(\{\X_i\}_{i\in T}\) depend. The key observation is that \emph{every} bit in \(\X\) depends on \emph{some} bit in \(\{\Y_i\}_{i\in T^\pr}\), by the maximality of \(T\). Using this observation, there are \emph{two possible cases}, over which we perform a \emph{win-win analysis}.

First, it is possible that \(T\) contains \(\geq k^\pr\) bits. In this case, we consider fixing all bits \(\{\Y_i\}_{i\notin T^\pr}\). It is then not too hard to show that \(\X\) becomes a source which contains \(\geq k^\pr\) good bits (which are mutually independent and not constants), and the remaining bad bits in \(\X\) are deterministic \(d\)-local functions of these good bits.\footnote{Technically, we need to fix a little more randomness to make this happen, but this can be done without much trouble by invoking some standard tricks from the extractor literature.} Thus in this case, we get that \(\X\) is a convex combination of NOBF sources of the desired type.

Second, it is possible that \(T\) contains \(<k^\pr\) bits. In this case, we consider fixing all bits \(\{\Y_i\}_{i\in T^\pr}\). But since all bits in \(\X\) depend on \emph{some} bit in this set, this fixing \emph{decrements} the locality \(d\to d-1\). And furthermore, since this fixes \(|T^\pr|\leq d|T|<dk^\pr\) bits, the entropy only decreases from \(k\to k-dk^\pr\) by the entropy chain rule. We then recurse until we hit the first case, or until we hit \(d=1\). If we eventually hit the first case, we already know that \(\X\) is a convex combination of NOBF sources of the desired type. On the other hand, it is easy to show that a \(1\)-local source is actually a \(1\)-local NOBF source! Thus we will always arrive at a (biased) \(d^\pr\)-local NOBF source with \(d^\pr\leq d\), proving that \(\X\) is always convex combination of NOBF sources of the desired type. Depending on when this recursion stops, we will arrive at an NOBF source with the number of good bits equal to at least
\[
\min\{k^\pr, k-dk^\pr,k-d(d-1)k^\pr,\dots,k-k^\pr\prod_{i\in[d]}i\}\geq\min\{k^\pr,k-d^2k^\pr\},
\]
which is always at least \(k^\pr\) provided \(k^\pr\leq \frac{k}{2d^2}\).

Thus we see that any \(d\)-local source with min-entropy \(k\) can be written as a convex combination of \(d\)-local NOBF source with \(\Omega(k/d^2)\) good bits, where the good bits are mutually independent (and nonconstant), but they may be heavily biased. So all that remains is to show that such biased \(d\)-local NOBF sources can be written as a convex combination of \emph{unbiased} \(d\)-local NOBF sources (as they were originally defined). This step is not difficult, by applying a standard Chernoff bound. However, since each good bit depends on up to \(d\) bits, each such good bit \(\X_i\) may have \(|\bias(\X_i)|=1-2\cdot2^{-d}\). As a result, we end up with \(\Omega(\frac{k}{d^22^d})\) unbiased good bits.

This completes the reduction from local to local NOBF sources. Given our earlier proof sketch that low-degree polynomials extract from local NOBF sources, we finally get that low-degree polynomials also extract from local sources, as desired.

\subsubsection{Low-degree dispersers for local sources}
We now proceed to sketch the proof of \cref{rem:main-remark-here}, which shows that most degree \(\leq r\) polynomials are dispersers for \(d\)-local sources with min-entropy at least
\[
k=O(d^3r\cdot(n\log n)^{1/r}).
\]
This improves our min-entropy requirement for extractors (which was \(k=O(2^dd^2r\cdot(2^d n\log n)^{1/r})\)) by removing two terms of the form \(2^d\). We use a different key idea to remove each \(2^d\) term. While the outer exponential term \(2^d\) is the more dramatic one to remove, it turns out that it is also the easier one. To do this, we simply note that for dispersers, we can forego the last step in our local to local NOBF reduction, which incurs a factor of \(2^d\) by making the biased local NOBF source into an unbiased one. This improves the entropy requirement for dispersers from
\[
k=O(2^dd^2r\cdot(2^d n\log n)^{1/r})\to k=O(d^2r\cdot(2^d n\log n)^{1/r}).
\]

\paragraph{Removing the inner exponential term.} Next, we focus on improving the entropy requirement for dispersers from
\[
k=O(d^2r\cdot(2^d n\log n)^{1/r})\to k=O(d^3r\cdot(n\log n)^{1/r}),
\]
turning the inner exponential term \(2^d\) into an outer linear term \(d\). This improvement is more challenging: while our first improvement relied on improving the local to local-NOBF reduction, this improvement relies on improving the entropy requirement for dispersing from local NOBF sources.

In order to show that low-degree polynomials extract from local NOBF sources, recall that we: (1) showed that a random low-degree polynomial extracts from an arbitrary local NOBF source with extremely high probability; and (2) used a union bound over the family \(\mathcal{X}\) of local NOBF sources to conclude that it extracts from \emph{all} local NOBF sources with high probability. To get our second improvement on the min-entropy requirement for dispersers, we get improved upper bounds on the size of \(\mathcal{X}\), so that our union bound is over fewer terms.

Towards this end, the key new idea is to show that in order to disperse from the family of \(d\)-local NOBF sources \(\mathcal{X}\), it actually suffices to disperse from the much smaller family \(\mathcal{X}^\pr\) of so-called \emph{\(d\)-local, degree \(\leq r\) NOBF sources}.\footnote{Technically, we require something slightly stronger than dispersion from such sources, but this does not make a huge difference. We will go into more details below.} This source family is the exact same as \(d\)-local NOBF sources, except it has the added restriction that the bad bits (which still depend on \(\leq d\) good bits each) can each be written as a degree \(\leq r\) polynomials. However, this family is significantly smaller: natural estimates on the sizes of \(\mathcal{X},\mathcal{X}^\pr\) give
\begin{align*}
    |\mathcal{X}|&\leq{n\choose k}\cdot\left({k\choose d}\cdot 2^{2^d}\right)^{n-k},\\
    |\mathcal{X}^\pr|&\leq{n\choose k}\cdot\left({k\choose d}\cdot2^{{d\choose \leq r}}\right)^{n-k}.
\end{align*}
After plugging in these improved size bounds, it is straightforward calculation to see that the inner \(2^d\) term from the entropy requirement drops out. So all that remains is to show the above claim that a disperser for \(d\)-local, degree \(\leq r\) NOBF sources automatically works for the more general family of \(d\)-local NOBF sources.

The key ingredient that goes into this claim is a simple lemma on polynomial decomposition. We show the following: for any function \(f:\zo^n\to\zo\), any degree \(\leq r\) polynomials \(a_1,\dots,a_n:\zo^k\to\zo\), and any polynomials \(b_1,\dots,b_n:\zo^k\to\zo\) that have no monomials of size \(\leq r\), the following holds. There exists a polynomial \(h:\zo^k\to\zo\) with no monomials of size \(\leq r\) such that
\[
f\left(a_1(y)+b_1(y), \dots, a_n(y)+b_n(y)\right)=f\left(a_1(y),\dots,a_n(y)\right)+h(y).
\]
Then, given an arbitrary \(d\)-local NOBF source \(\X\sim\zo^n\), the idea is to write it in the form
\[
\X=(a_1(\Y)+b_1(\Y),\dots,a_n(\Y)+b_n(\Y)),
\]
where \(\Y\sim\zo^k\) is uniform and \(a_i,b_i\) are as before. Using our polynomial decomposition lemma, it then (roughly) holds that \(f\) is a disperser for \(\X\) if \(f\) is a disperser for the simpler class of \(d\)-local, degree \(\leq r\) NOBF sources. More precisely, we actually require from \(f\) a property that is ever-so-slightly stronger than being a disperser: we require that for any \(d\)-local, degree \(\leq r\) NOBF source \(\X^\pr=(a_1(\Y),\dots,a_n(\Y))\), it holds that the polynomial \(f(a_1,\dots,a_n)\) has a monomial of degree \(\leq r\). Our polynomial decomposition lemma then guarantees that \(f\) will also have this property for the more general \(d\)-local source, since the polynomial \(h\) in our decomposition lemma does not have any monomials of degree \(\leq r\). Intuitively, \(h\) is not able to ``destroy'' the monomial of degree \(\leq r\) guaranteed to pop out of \(f(a_1(y),\dots,a_n(y))\).

Thus, it suffices to ``disperse'' from \(d\)-local, degree \(\leq r\) NOBF sources in order to disperse from more general \(d\)-local NOBF sources, meaning that we can leverage our improved bound on the size of \(\mathcal{X}^\pr\) to get our claimed improvement on the disperser's entropy requirement.

\subsection{Lower bounds}

We now discuss our entropy lower bounds for low-degree extraction from local sources. By this, we mean that for every degree \(\leq r\) polynomial \(f\in\F_2[x_1,\dots,x_n]\) we can find a \(d\)-local source \(\X\sim\zo^n\) with relatively high min-entropy \(k\) on which \(f\) is constant. In other words, we show that in order to disperse (and thus extract) from \(d\)-local sources, they must have min-entropy exceeding this value \(k\).

We show that every degree \(r\leq\Omega(\log n)\) polynomial \(f\in\F_2[x_1,\dots,x_n]\) must admit a \(d\)-local source \(\X\sim\zo^n\) of min-entropy at least
\[k=\Omega(r(dn\log{n})^{1/r})\]
on which it is constant. That is, we sketch the proof of our lower bound (\cref{thm:localtightintro}).

In order to prove this result, we actually prove a slightly stronger result: we show that we can find a \(d\)-local source \(\X\sim\zo^n\) with the above parameters such that it is also \emph{affine}.

Our starting point is a tight result of Cohen and Tal \cite{CT15}, which shows that any degree \(\leq r\) polynomial \(f:\F_2^n\to\F_2\) admits a subspace \(V\subseteq\F_2^n\) of dimension \(\Omega(rn^{1/(r-1)})\) on which it is constant. Here, we obtain a (tight) ``local'' version of their result, and show that any degree \(\leq r\) polynomial \(f\) admits a \(d\)-local subspace \(X\subseteq\F_2^n\) of dimension \(k=\Omega(r(dn\log n)^{1/r})\) on which it is constant. Here, we say that \(V\) is \emph{\(d\)-local} if \(V\) has a basis \(v_1,\dots,v_k\in\F_2^n\) such that for any index \(i\in[n]\), at most \(d\) of these basis vectors equal \(1\) at this index. It is straightforward to verify that the uniform distribution \(\X\) over \(V\) is a \(d\)-local source with min-entropy \(k\), so we focus now on proving the existence of such a \(V\).

At a high level, the proof of Cohen and Tal proceeds by iteratively growing a subspace \(V\) on which \(f\) is constant. At each phase, they define a set \(A\subseteq\F_2^n\) such that \(f\) is constant over \(\mathsf{span}(V,x)\) for every \(x\in A\). They note that if \(|A|\) has size \(>2^{\mathsf{dim}(V)}\), then of course there is some \(x\in A\setminus V\) and furthermore we already know that \(f\) is constant on \(\mathsf{span}(V,x)\). Thus, they can grow their monochromatic subspace by one dimension.

In order to get a lower bound on \(|A|\), they note that this set can be defined as the common solutions to a small collection of low-degree polynomials. A classical result known as the \emph{Chevalley-Warning theorem} (\cref{thm:absolutely-standard-Chevalley-Warning}) then shows that \(|A|\geq2^{n-t}\), where \(t\) is the sum of degrees across the collection of polynomials. To complete their proof, they grow their subspace \(V\) until they are no longer able to show \(|A|>2^{\mathsf{dim}(V)}\).

In our lower bound, we show that \(f\) is monochromatic on a \emph{\(d\)-local subspace}. To prove this, we start with the same approach as Cohen and Tal. However, at each phase, we add extra constraints to \(A\) which guarantee the following: if we take any \(x\in A\) and add it to our current subspace \(V\) (with basis, say, \(v_1,\dots,v_{|\dim(V)|}\)), then the location of the \(1\)s appearing in vectors \(x,v_1,\dots,v_{|\dim(V)|}\) satisfy the \(d\)-locality constraint defined above. Again, as long as \(A\) is large enough, we can find some \(x\in A\) that grows the dimension of our \(d\)-local subspace.

In order to ensure that \(A\) remains large for as many iterations as possible, we would like to minimize the impact of the new ``locality'' constraints that we have added to \(A\). Given the description of these constraints above, we observe that these constraints are minimized if we grow \(V\) by carefully selecting vectors that have the lowest possible Hamming weight. However, we now need an upper bound on the Hamming weight of the lightest (nontrivial) common solution to a system of polynomial equations. Thus, our key new ingredient will be a result of this type, which we call a ``low-weight Chevalley-Warning theorem.''

\paragraph*{A low-weight Chevalley-Warning theorem.}
Above, we saw how the classical Chevalley-Warning theorem is critical in lower bounding the size of \(A\), thereby showing that there is some (nontrivial) vector \(v\in A\) by which we can grow our monochromatic subspace. Now, we need an additional guarantee that there is such a \(v\in A\) that also has low Hamming weight. We prove such a result, and call it a \emph{low-weight Chevalley-Warning theorem}. Our theorem roughly says the following. Given a collection \(\{f_i\}\) of polynomials that have cumulative degree \(D\) (and a common solution \(0\)), if most of these polynomials have degree \(\leq 1\) then they admit a nontrivial solution of Hamming weight at most
\[
w=O(D/\log(n/D)).
\]

In order to prove our result, our key observation is that for any large enough subset \(A\) of common solutions to \(\{f_i\}\), it holds that \(A+A\) also contains a (nontrivial) common solution to \(\{f_i\}\). As it turns out, this observation follows quite readily from the CLP lemma \cite{CLP17} - a result which was instrumental in the recent resolution of the cap set conjecture. Furthermore, given the above observation, we obtain an elementary proof of our low-weight Chevalley-Warning theorem, as follows. First, we let \(Q\) denote the set of common solutions to our collection of polynomials, and then:
\begin{enumerate}
    \item We assume for contradiction that there is no nontrivial solution \(q\in Q\) of weight \(\leq w\).
    \item We use our key observation to conclude that every Hamming ball of radius \(w/2\) cannot have too many elements of \(Q\) in it, which implies that \(Q\) is a list-decodable code with small list size at radius \(w/2\).
    \item Using the list-decoding properties of \(Q\), we use the Hamming bound (for list-decodable codes) to get an \emph{upper bound} on its size.
    \item Using the fact that \(Q\) holds the common solutions to a small set of low-degree polynomials, we use the classical Chevalley-Warning theorem to get a \emph{lower bound} on its size.
    \item We observe that the lower bound is greater than the upper bound, which yields a contradiction.
\end{enumerate}

Equipped with our low-weight Chevalley-Warning theorem, our entropy lower bound for low-degree extraction from \(d\)-local affine spaces follows immediately via the proof sketch described above.

\subsection{A barrier}
To conclude our overview, we provide a proof sketch of our barrier result (\cref{claim:barrierintro}), which shows that affine extractors (applied in a black-box manner) cannot extract from local sources with min-entropy \(k=\Omega(\sqrt{n})\), even if the locality is \(2\). More formally, to show that an affine extractor also extracts from a different family \(\mathcal{Q}\) of distributions with min-entropy \(k\), the standard technique is to show that each source \(\Q\in\mathcal{Q}\) with min-entropy \(k\) is (close to) a convex combination of affine sources with min-entropy slightly less than \(k\). Here, we show that this is simply not possible for local sources with min-entropy \(\sqrt{n}\). In particular, we show that there is a very simple \(2\)-local source \(\Q\sim\zo^n\) with min-entropy \(\Omega(\sqrt{n})\) that has statistical distance \emph{exponentially close to \(1\)} from any convex combination of affine sources with min-entropy \(k^\pr\).

In more detail, we consider the \(2\)-local ``clique'' source \(\Q\sim\zo^n\) defined as follows: first, pick any \(\ell\in\N\) and set \(n:=\ell+{\ell\choose 2}\). Then, pick uniform and independent bits \(\mathbf{q}_1,\dots,\mathbf{q}_\ell\sim\zo\) and set \(\Q\) to be the concatenation of all \(\mathbf{q}_i\) over \(1 \le i \le n\) and \(\mathbf{q}_i\cdot\mathbf{q}_j\) over \(1\leq i<j\leq \ell\). Now, let \(\X\sim\zo^n\) be a convex combination of affine sources, each with min-entropy \(\ell^\pr\). We argue that \(|\Q-\X|\geq1-2^{-\Omega(\ell^\pr)}\) by showing that for any \(\ell^\pr\)-dimensional affine \(\F_2\)-subspace \(S\subseteq\zo^n\), it holds that \(|\supp(\Q)\cap S|/|S|\leq 2^{-\Omega(\ell^\pr)}\). That is, we wish to show that \(Q=\supp(\Q)\) is \emph{subspace-evasive}.

To show that cliques are subspace-evasive, we use the following key observation:
For any nonempty set \(Q^\pr\subseteq Q\) of  cliques, the set \(Q^\pr + Q^\pr := \{u+v : u\in Q^\pr,v\in Q^\pr, u\neq v\}\) (where the sum is over \(\F_2^n\)) has a ``Sidon property:''  each element \(x\) in \(Q^\pr + Q^\pr\) has a unique pair \(u,v \in Q^\pr\) such that \(x = u+v\). This observation is proven by noticing that by making another copy of each coordinate of the form \(\mathbf{q}_i\cdot\mathbf{q}_j\), the set \(Q\) will correspond precisely to the symmetric rank-1 matrices of \(\F_2^{\ell \times \ell}\). Thus all the elements in \(Q^\pr + Q^\pr\) would correspond to symmetric \(\F_2^{\ell \times \ell}\) matrices of rank at most $2$. Hence by looking at the row space of \(x \in Q^\pr + Q^\pr\), we can precisely find its symmetric rank-1 decomposition. That is, we can find \(u,v \in Q^\pr\) such that $x=u+v$. Now, pick \(Q^\pr = Q \cap S\). Since addition is closed in \(S\), we see that \(Q^\pr + Q^\pr \subseteq S\). Thus \(|S| \geq |Q^\pr + Q^\pr| \geq {|Q^\pr|\choose 2} \geq \Omega(|Q\cap S|^2)\). Hence we find that \(|Q\cap S|/|S|\leq O(\sqrt{|S|}/|S|)=O(1/\sqrt{|S|})\), which is \(2^{-\Omega(\ell^\pr)}\) as \(|S| = 2^{\ell^\pr}\).

\section{Preliminaries}\label{sec:new-preliminaries}

We outline some basic notation, definitions, and facts that will be used throughout the paper.

\subsection{Notation}
We let \(\N:=\{1,2,\dots,\}\) denote the natural numbers, and for any \(n\in\N\), we define \([n]:=\{1,2,\dots,n\}\). Throughout, we use \(\log\) to denote the base-\(2\) logarithm, and we write \(\binom{n}{\leq r}:=\sum_{i=0}^r\binom{n}{i}\). For a bitstring \(x\in\zo^n\), we let \(x_i\in\zo\) denote the bit it holds at coordinate \(i\), and for a set \(S\subseteq[n]\) we let \(x_S\in\zo^{|S|}\) denote the the concatenation of bits \(x_i,i\in S\) in increasing order of \(i\). On the other hand, we define \(x^S:=\prod_{i\in S}x_i\), and let \(x^\emptyset:=1\).

We define the \emph{support} of \(x\in\zo^n\) as the set of coordinates \(i\) where \(x_i=1\).

For any \(n\in\N\), we let \(0^n\) denote the string of \(n\) \(0\)s in a row, and we let \(1^n\) denote the string of \(n\) \(1\)s in a row. Given a set \(S\subseteq[n]\), we write \(\overline{S}\) to denote its complement. We let \(\F_2\) denote the finite field of size \(2\), and we let \(\F_2^n\) denote the \(n\)-dimensional vector space over \(\F_2\). Given sets \(A,B\subseteq\F_2^n\), we let \(A+B\) denote the \emph{sumset} of \(A\) and \(B\):
\[
A+B:=\{a+b : a\in A, b\in B\}.
\]

\subsection{Probability theory}
We now overview some basic notions from (discrete) probability theory. First, random variables are denoted by boldface letters such as \(\X\), and we let \(\supp(\X)\) denote its support. For a set \(S\), we write \(\X\sim S\) to denote that \(\X\) is supported on a subset of \(S\). Next, the uniform distribution over \(\zo^m\) is denoted by \(\U_m\), and whenever \(\U_m\) appears multiple times in the same expression or formula, this represents multiple copies of the \emph{same} random variable. For example, the random variable \((\U_m,\U_m)\) hits each string \((x,x)\in\zo^m\times\zo^m\) with probability \(2^{-m}\).

Throughout, we will measure ``randomness'' of a random variable by its min-entropy:

\begin{definition}[Min-entropy]\label{def:minent}
The \emph{min-entropy} of a random variable $\X$ supported on a set $S$, denoted by $\minH(\X)$, is defined as
\begin{equation*}
    \minH(\X)=-\log \max_{x\in S}\Pr[\X=x].
\end{equation*}
\end{definition}

The following lemma about min-entropy is not difficult to show, but is extremely useful.

\begin{lemma}\label{lem:clean-entropy-drop}
Suppose $\X$ and $\Y$ are arbitrary random variables such that $\Y$ is uniformly distributed over its support.
Then, for every \(y\in\supp(\Y)\), it holds that
\begin{equation*}
    \minH(\X|\Y=y)\geq\minH(\X)-\log|\supp(\Y)|.
\end{equation*}
\end{lemma}

We measure the similarity of two random variables (or, rather, their distributions) via statistical distance.

\begin{definition}[Statistical distance]
The \emph{statistical distance} between two random variables $\X$ and $\Y$ supported on a set $S$, denoted by $\Delta(\X,\Y)$, is defined as
\begin{equation*}
    \Delta(\X,\Y)=\max_{T\subseteq S}|\Pr[\X\in T]-\Pr[\Y\in T]|=\frac{1}{2}\sum_{s\in S}|\Pr[\X=s]-\Pr[\Y=s]|.
\end{equation*}
Moreover, we say $\X,\Y$ are \emph{$\eps$-close}, denoted $\X\approx_\eps \Y$, if $\Delta(\X,\Y)\leq \eps$.

\noindent On the other hand, if \(\Delta(\X,\Y)\geq\eps\), we say that \(\X,\Y\) are \emph{\(\eps\)-far}.
\end{definition}

Next, we say that \(\X\) is a \emph{convex combination} of distributions \(\{\Y_i\}\) if there exist probabilities \(\{p_i\}\) summing to \(1\) such that \(\X=\sum_i p_i\Y_i\), meaning that \(\X\) samples from \(\Y_i\) with probability \(p_i\). The following standard fact about convex combinations is extremely useful in extractor research, and is straightforward to show.

\begin{fact}\label{fact:convex-combo-lifted-extractor}
Let \(\Ext:\zo^n\to\zo\) be an extractor for a family of distributions \(\mathcal{X}\) over \(\zo^n\) with error \(\eps\). Then \(\Ext\) also extracts with error \(\eps\) from any \(\X\) which is a \emph{convex combination} of distributions from \(\mathcal{X}\).
\end{fact}

For the easier setting of dispersers, it is easy to show something even stronger than the above fact.

\begin{fact}\label{fact:convex-combo-lifted-disperser-optimized}
Let \(\Disp:\zo^n\to\zo\) be a disperser for a family of distributions \(\mathcal{X}\) over \(\zo^n\). Then \(\Disp\) is also a disperser for any \(\X\) whose support contains the support of some \(\X^\pr\in\mathcal{X}\).
\end{fact}

Finally, we will make use of the following standard concentration bound.

\begin{lemma}[Chernoff bound, lower tail]\label{lem:chernoff}
Let \(\X_1,\dots,\X_n\) be independent random variables over \(\zo\), and define \(\mathbf{Z}:=\sum_{i\in[n]}\X_i\). Then for every \(0<\delta<1\), it holds that
\[
\Pr[\mathbf{Z}\leq(1-\delta)\cdot\E[\mathbf{Z}]]\leq e^{-\E[\mathbf{Z}]\cdot\delta^2/2}.
\]
\end{lemma}

\subsection{Coding theory}

We will now review some basics of coding theory, which will be used in both our entropy upper bounds and lower bounds. First, given a string \(v\in\zo^n\), we let \(\Delta(v):=\{i\in[n] : v_i\}\) denote its \emph{Hamming weight}, and for any strings \(u,v\in\zo^n\), we let \(\Delta(u,v):=\{i\in[n] : u_i\neq v_i\}\) denote their \emph{Hamming distance}. Then, given any string \(v\in\zo^n\) and integer \(r\in\{0,1,\dots,n\}\), we let
\[
\mathcal{B}(v,r):=\{v^\pr\in\zo^n : \Delta(v,v^\pr)\leq r\}
\]
denote the (closed) Hamming ball of radius \(r\) around (i.e., centered at) \(v\). If we replace the inequality ``\(\leq\)'' above with a strict inequality ``\(<\)'', we call this object the \emph{open} Hamming ball of radius \(r\) around \(v\), and denote it by \(\mathcal{B}^-(v,r)\). Notice that the closed Hamming ball has size \(\binom{n}{\leq r}\), whereas the open Hamming ball has size \(\binom{n}{\leq r-1}\).

An \((n,k,d)\) code is a subset \(Q\subseteq\zo^n\) of size \(2^k\) such that the pairwise Hamming distance between any distinct \(x,y\in Q\) is at least \(d\). We call an \((n,k,d)\) code \(Q\subseteq\zo^n\) \emph{linear} if it is an \(\F_2\)-linear subspace, and subsequently call it an \([n,k,d]\) code.

A standard relaxation of \((n,k,d)\) codes are \emph{list-decodable codes}. We say that a subset \(Q\subseteq\zo^n\) is a \((\rho,L)\) list-decodable code if every Hamming ball of radius at most \(\rho n\) contains at most \(L\) codewords (i.e., elements of \(Q\)). The following classical bound places a limit on the tradeoff between the size and list-decodability of arbitrary codes.

\begin{theorem}[Hamming bound for list-decodable codes]\label{thm:hamming-bound}
For any \((\rho,L)\)-list-decodable code \(Q\subseteq\zo^n\),
\[
|Q|\leq\frac{2^n L}{\binom{n}{\leq \rho n}}.
\]
\end{theorem}
\begin{proof}
Consider the quantity \(\sum_{q\in Q}|\mathcal{B}(q,\rho n)|\). Notice that for any fixed \(v\in\zo^n\), there are at most \(L\) codewords \(q\in Q\) such that \(v\in\mathcal{B}(q,\rho n)\), because otherwise \(|\mathcal{B}(v,\rho n)\cap Q|>L\), contradicting the list-decodability of \(Q\). Thus
\[
|Q|\cdot\binom{n}{\leq\rho n}=\sum_{q\in Q}|\mathcal{B}(q,\rho n)|\leq\sum_{v\in\zo^n}L\leq2^nL,
\]
and the inequality follows.
\end{proof}

Finally, a classic linear code that will be used in this paper is the \emph{Reed-Muller code}, defined below.

\begin{definition}
The \emph{Reed-Muller code} \(\mathsf{RM}(m,r)\) is the subset \(Q\subseteq\F_2^{2^m}\) defined as follows:
\[
Q:=\{ (p(\alpha))_{\alpha\in\F_2^m} : p\text{ is a multilinear polynomial of degree \(\leq r\).}\}
\]
It is a \([2^m,\binom{m}{\leq r},2^{m-r}]\) linear code.
\end{definition}

\subsection{Boolean functions and \(\F_2\)-polynomials}\label{subsec:prelims:polynomials}

We now review some standard definitions and facts about boolean functions and \(\F_2\)-polynomials, which will be used throughout the paper. First, every function \(f:\F_2^n\to\F_2\) has a unique representation as a multilinear \(\F_2\)-polynomial. This means that there exist (unique) coefficients \(\{c_S\}_{S\subseteq[n]}\), each from \(\F_2\), so that
\[
f(x) = \sum_{S\subseteq[n]}c_Sx^S.
\]
The \emph{degree} of \(f\), denoted \(\deg(f)\), is the size of the largest \(S\) such that \(c_S=1\), and we say that \(\deg(f)=-\infty\) if all \(c_S=0\). A \emph{random degree \(\leq r\) polynomial} (independently) sets each \(c_S\) to \(1\) with probability \(1/2\) if \(|S|\leq r\), and otherwise sets \(c_S=0\). We often do not distinguish between a function \(f:\zo^n\to\zo\) and its natural interpretation as a function \(f:\F_2^n\to\F_2\), nor do we distinguish between a function \(f:\F_2^n\to\F_2\) and its unique representation as a multilinear \(\F_2\)-polynomial. This allows us to slightly abuse language, and gives meaning to phrases like ``the degree of \(f:\zo^n\to\zo\).''

Next, we let \(\F_2[x_1,\dots,x_n]\) denote the set of (multilinear) \(\F_2\)-polynomials, and we let \(\{f_i\}\subseteq\F_2[x_1,\dots,x_n]\) refer to a set (of unspecified size) of such polynomials. The \emph{linear degree} of \(\{f_i\}\) is the sum of the degrees of the \(f_i\)'s which have degree \(1\), the \emph{nonlinear degree} of \(\{f_i\}\) is the sum of the degrees of the \(f_i\)'s which have degree \(>1\), and the \emph{degree} of \(\{f_i\}\) is the sum of the degrees of all the \(f_i\)'s. We say that \(x\) is a \emph{common solution} to \(\{f_i\}\) if \(f_i(x)=0\) for all \(i\), and we say that \(x\) is \emph{nontrivial} if \(x\neq 0\).

The notion of \emph{bias} and \emph{correlation} of boolean functions will be important for our entropy upper bounds. Given arbitrary functions \(f,g:\zo^n\to\zo\), we define their \emph{correlation} as
\[
\corr(f,g):=|\E_{x\sim\zo^n}[(-1)^{f(x)+g(x)}]|=|\Pr[f(x)=g(x)]-\Pr[f(x)\neq g(x)]|,
\]
and we define the \emph{bias} of \(f\) as
\[
\bias(f):=|\E_{x\sim\zo^n}[(-1)^{f(x)}]|=|\Pr[f(x)=0]-\Pr[f(x)=1]|.
\]
Observe that \(\bias(f)=\corr(f,0)\).

To conclude this section, we record some notation about function combination and composition. For any two functions \(a,b:\F_2^k\to\F_2\), we let \(a+b:\F_2^k\to\F_2\) denote the function \((a+b)(x):=a(x)+b(x)\), and we let \(ab:\F_2^k\to\F_2\) denote the function \(ab(x):=a(x)b(x)\). For any functions \(f:\F_2^n\to\F_2\) and \(a_1,\dots,a_n:\F_2^k\to\F_2\), we let \(f(a_1,\dots,a_n):\F_2^k\to\F_2\) denote the function
\(
f(a_1,\dots,a_n)(x):=f(a_1(x),\dots,a_n(x))
\).
Finally, given functions \(a_1,\dots,a_n:\F_2^k\to\F_2\) and \(S\subseteq[n]\), we defined \(a^S:\F_2^k\to\F_2\) as \(a^S:=\prod_{i\in S}a_i\), and we let \(a^\emptyset:=1\).

\section{Entropy upper bounds}\label{sec:new-upper-bounds}

In this section, we obtain upper bounds on the entropy required to extract from \(d\)-local sources using degree \(\leq r\) polynomials, proving the following theorem.

\begin{theorem}[\cref{thm:localextintro}, restated]\label{thm:localextintro:restated}
There are universal constants \(C,c>0\) such that for all \(n,d,r\in\N\), the following holds. With probability at least \(0.99\) over the choice of a random degree \(\leq r\) polynomial \(f\in\F_2[x_1,\dots,x_n]\), it holds that \(f\) is an \(\eps\)-extractor for \(d\)-local sources with min-entropy \(\geq k\) and error \(\eps=2^{-\frac{ck}{r^32^dd^2}}\), as long as
\[
k\geq C2^dd^2r\cdot(2^dn\log n)^{1/r}.
\]
\end{theorem}

In order to prove \cref{thm:localextintro:restated}, we combine two key ingredients. Our first key ingredient reduces \(d\)-local sources to the more specialized family of \(d\)-local NOBF sources:

\begin{theorem}[\cref{thm:red-to-nobf}, restated]\label{thm:upper-bounds:first-key-ingredient}
There exists a universal constant \(c>0\) such that for any \(n,k,d\in\N\), the following holds. Let \(\X\sim\zo^n\) be a \(d\)-local source with min-entropy \(\geq k\). Then \(\X\) is \(\eps\)-close to a convex combination of \(d\)-local NOBF sources with min-entropy \(\geq k^\pr\), where \(\eps=2^{-ck^\pr}\) and
\[
k^\pr=\frac{ck}{2^d d^2}.
\]
\end{theorem}

Our second key ingredient gives upper bounds on the entropy required to extract from \(d\)-local NOBF sources using degree \(\leq r\) polynomials:

\begin{theorem}\label{thm:upper-bounds:second-key-ingredient}
There are universal constants \(C,c>0\) such that for all \(n,d,r\in\N\), the following holds.
With probability at least \(0.99\) over the choice of a random degree \(\leq r\) polynomial \(f\in\F_2[x_1,\dots,x_n]\), it holds that \(f\) is an \(\eps\)-extractor for \(d\)-local NOBF sources of min-entropy \(\geq k\) with error \(\eps=2^{-ck/r^3}\), as long as
\[
k\geq C r\cdot(2^d\cdot n\log n)^{1/r}.
\]
\end{theorem}

Recall that if an extractor works for a family \(\mathcal{X}\) of sources over \(\zo^n\), then it also works for any \(\X\sim\zo^n\) which is a convex combination of sources from \(\mathcal{X}\) (\cref{fact:convex-combo-lifted-extractor}). As a result, by combining \cref{thm:upper-bounds:first-key-ingredient,thm:upper-bounds:second-key-ingredient}, we immediately get \cref{thm:localextintro:restated}. Thus, in the remainder of this section, we just focus on proving \cref{thm:upper-bounds:first-key-ingredient,thm:upper-bounds:second-key-ingredient}. We prove \cref{thm:upper-bounds:first-key-ingredient} in \cref{subsec:upper-bounds:first-key-ingredient}, and we prove \cref{thm:upper-bounds:second-key-ingredient} in \cref{subsec:upper-bounds:second-key-ingredient}. We conclude the section in \cref{subsec:optimized-dispersers}, where we show that \cref{thm:localextintro:restated} can be optimized for the easier setting of \emph{dispersers}, allowing us to obtain a formal version of \cref{rem:main-remark-here}.

\subsection{A reduction from \(d\)-local sources to \(d\)-local NOBF sources}\label{subsec:upper-bounds:first-key-ingredient}

We start by proving our reduction, \cref{thm:upper-bounds:first-key-ingredient}. In order to reduce \(d\)-local sources to \(d\)-local NOBF sources, we use an intermediate model called \emph{biased \(d\)-local NOBF sources}.

\begin{definition}
A random variable \(\X\sim\zo^n\) is a \emph{\((\delta,k)\)-biased \(d\)-local NOBF source} if there exists a set \(S\subseteq[n]\) of size \(k\) such that both of the following hold:
\begin{itemize}
	\item The bits in \(\X_S\) are mutually independent (but need not be identically distributed), and each \(\X_i,i\in S\) has bias \(|\Pr[\X_i=1]-\Pr[\X_i=0]|\leq\delta\).
	\item Every other bit \(\X_j,j\notin S\) is a deterministic function of at most \(d\) bits in \(\X_S\).
\end{itemize}
\end{definition}

Notice that this intermediate model generalizes \(d\)-local NOBF sources of min-entropy \(k\), which are just \((0,k)\)-biased \(d\)-local NOBF sources. Now, given this intermediate model, we prove \cref{thm:upper-bounds:first-key-ingredient} by combining two lemmas.

The first lemma reduces \(d\)-local sources to biased \(d\)-local NOBF sources.

\begin{lemma}[Reduction, Part 1]\label{lem:hard-part-of-reduction}
Let \(\X\sim\zo^n\) be a \(d\)-local source with min-entropy \(\geq k\). Then \(\X\) is a convex combination of \((\delta,k^\pr)\)-biased \(d\)-local NOBF sources, where \(\delta\leq1-2^{-d}\) and \(k^\pr\geq k/(2d^2)\).
\end{lemma}

The second lemma reduces biased \(d\)-local NOBF sources to (unbiased) \(d\)-local NOBF sources.

\begin{lemma}[Reduction, Part 2]\label{lem:simulate-coins}
Let \(\X\sim\zo^n\) be a \((\delta,k)\)-biased \(d\)-local NOBF source. Then \(\X\) is \(\eps\)-close to a convex combination of \((0,k^\pr)\)-biased \(d\)-local NOBF sources, where \(k^\pr\geq(1-\delta)k/4\) and \(\eps=2^{-(1-\delta)k/4}\).
\end{lemma}

By combining these two lemmas, \cref{thm:upper-bounds:first-key-ingredient} follows immediately. Thus, all that remains in the proof of our reduction (\cref{thm:upper-bounds:first-key-ingredient}) is to prove \cref{lem:hard-part-of-reduction,lem:simulate-coins}, which we do in \cref{subsubsec:first-part-reduction,subsubsec:second-part-reduction}, respectively.

\subsubsection{A reduction from \(d\)-local sources to biased \(d\)-local NOBF sources}\label{subsubsec:first-part-reduction}

We now prove the first part of our reduction (\cref{lem:hard-part-of-reduction}), which shows that every \(d\)-local source is a convex combination of biased \(d\)-local NOBF sources.
\begin{proof}[Proof of \cref{lem:hard-part-of-reduction}]
Let \(\X\sim\zo^n\) be a \(d\)-local source with min-entropy \(\geq k\). We wish to show that \(\X\) is a convex combination of \((\delta,k^\pr)\)-biased \(d\)-local NOBF sources, where \(\delta\leq 1-2^{-d}\) and \(k^\pr\geq k/(2d^2)\).

The key observation that we will prove is that for any \(t\), one of the following \emph{must} hold: either
\begin{itemize}
	\item \(\X\) is a convex combination of \((\delta,t)\)-biased \(d\)-local NOBF sources, for \(\delta\leq1-2^{-d}\); or
	\item \(\X\) is a convex combination of \((d-1)\)-local sources with min-entropy \(>k-td\).
\end{itemize}
Before we prove this key observation, let us see how we can use it to prove the desired result. First, recall that convex combinations ``stack'' in the following sense: if a source \(\X\) is a convex combination of convex combinations of sources from a family \(\mathcal{X}\), then \(\X\) is just a convex combination of sources from \(\mathcal{X}\). Thus, by repeatedly applying the key observation until either the first item becomes true or we arrive at a \(1\)-local source (the ``base case''), we see that \(\X\) is a convex combination of sources \(\{\mathbf{Z}_i\}\), where each \(\mathbf{Z}_i\) is either:
\begin{itemize}
	\item A \((\delta,t)\)-biased \(d\)-local NOBF source, for \(\delta\leq1-2^{-d}\); or
	\item A \(1\)-local source with min-entropy \(>k - t\cdot(d + (d-1) + \dots + 2)=k-t\cdot(d^2+d-2)\).
\end{itemize}

However, it is clear from the definitions that a \(1\)-local source with min-entropy \(k^\pr\) is a \(1\)-local NOBF source with min-entropy \(k^\pr\). Furthermore, it is easy to see that a \(1\)-local NOBF source with min-entropy \(\geq k^\pr\) is a convex combination of \(1\)-local NOBF sources with min-entropy exactly \(k^\pr\), by fixing any additional random ``good'' bits. Thus, for any \(t\leq k^\pr\), we know that a \(1\)-local source with min-entropy \(\geq k^\pr\) is a convex combination of \((\delta,t)\)-biased \(d\)-local NOBF sources, for \(\delta\leq1-2^{-d}\).

By the above discussion, we see that for any \(t\leq k-t\cdot(d^2+d-2)\), \(\X\) is a convex combination of \((\delta,t)\)-biased \(d\)-local NOBF sources, where \(\delta\leq1-2^{-d}\). Setting \(t=\frac{k}{2d^2}\) yields the result.

Thus, all that remains is to prove the key observation stated at the beginning of the proof. Towards this end, let \(\X\sim\zo^n\) be a \(d\)-local source with min-entropy \(\geq k\). By definition of \(d\)-local source, there exists some \(\ell\) and \(f:\zo^\ell\to\zo^n\) such that \(\X=f(\Y)\) for uniform \(\Y\sim\U_\ell\), such that each bit \(\X_i\) is a deterministic function of at most \(d\) bits in \(\Y\). In other words, there exist sets \(S_1,\dots,S_n\subseteq[\ell]\) of size \(d\) and functions \(f_1,\dots,f_n:\zo^d\to\zo^n\) such that
\[
\X=(\X_1,\X_2,\dots,\X_n)=(f_1(\Y_{S_1}),f_2(\Y_{S_2}),\dots,f_n(\Y_{S_n})).
\]
Now, let \(T\subseteq[n]\) be any set of coordinates of \emph{maximal size} such that:
\begin{itemize}
\item \(H_\infty(\X_i)>0\) for all \(i\in T\); and
\item \(S_i\cap S_j=\emptyset\) for any distinct \(i,j\in T\).
\end{itemize}
Suppose \(T\) has size \(\tau\). Without loss of generality, assume \(T=[\tau]\). We conclude with two cases.

\textbf{Case (i)}: \(\tau<t\). In this case, we fix the random variable \(\Y_{S_1},\dots,\Y_{S_\tau}\). We know that with probability \(1\) over this fixing, all bits \(\X_i,i\in[n]\) become deterministic functions of at most \(d-1\) unfixed variables in \(\Y\), by the maximality of \(T\) and its intersection property. In other words, \(\X\) becomes a \((d-1)\)-local source. Furthermore, by \cref{lem:clean-entropy-drop}, we know that with probability \(1\) over this fixing, \(\X\) loses \(\sum_{i\in[\tau]}|S_i|=d\tau<dt\) bits of min-entropy. Thus in this case, \(\X\) is a convex combination of \((d-1)\)-local sources of min-entropy \(>k-dt\).

\textbf{Case (ii)}: \(\tau\geq t\). In this case, define \(\overline{S}:=[n]-(\bigcup_{i\in[\tau]}S_i)\) and notice that \(S_1,S_2,\dots,S_\tau,\overline{S}\) partition the coordinates of \(\Y\). Next, define the random variables \(\mathbf{Z}_i:=\Y_{S_i}\) for each \(i\in[\tau]\), and define \(\overline{\mathbf{Z}}:=\Y_{\overline{S}}\). Notice that \(\X_i=f_i(\mathbf{Z}_i)\) for each \(i\in[\tau]\). Furthermore, it is straightforward to verify that for all \(j>\tau\), there exists a set \(Q_j\subseteq[\tau]\) of size at most \(d\) and a deterministic function \(f^\pr_j\) such that \(\X_j=f^\pr_j(\mathbf{Z}_{Q_j},\overline{\mathbf{Z}})\). In other words, we can rewrite \(\X\) as
\begin{align*}
\X&=(\X_1,\dots,\X_{\tau},\X_{\tau+1},\dots,\X_n)\\
&=(f_1(\mathbf{Z}_1),\dots,f_\tau(\mathbf{Z}_\tau),f_{\tau+1}^\pr(\mathbf{Z}_{Q_{\tau+1}},\overline{\mathbf{Z}}),\dots,f_n^\pr(\mathbf{Z}_{Q_n},\overline{\mathbf{Z}})).
\end{align*}
Now, for each \(i\in[\tau]\), define \(\mathbf{A}_i:=f_i(\mathbf{Z}_i)\). Furthermore, it is straightforward to show that we can define a new random variable \(\mathbf{B}\) independent of \(\Y\), and for each \(i\in[\tau]\) a deterministic function \(g_i\) such that \(g_i(\mathbf{A}_i,\mathbf{B})=\mathbf{Z}_i\) for all \(i\in[\tau]\). Thus, for any subset \(Q\subseteq[\tau]\) we have \(\mathbf{Z}_Q=g^\pr_Q(\mathbf{A}_Q,\mathbf{B})\) for some deterministic function \(g^\pr_Q\). And finally, for each \(j>\tau\) there must be some deterministic function \(\psi_j\) such that
\[
f^\pr_{j}(\mathbf{Z}_{Q_j},\overline{\mathbf{Z}})=\psi_j(\mathbf{A}_{Q_j},\mathbf{B},\overline{\mathbf{Z}}).
\]

Thus we can rewrite \(\X\) as:
\[
\X=(\mathbf{A}_1,\dots,\mathbf{A}_\tau,\psi_{\tau+1}(\mathbf{A}_{Q_{\tau+1}},\mathbf{B},\overline{\mathbf{Z}}),\dots,\psi_n(\mathbf{A}_{Q_n},\mathbf{B},\overline{\mathbf{Z}})).
\]
Notice that the collection \(\{\mathbf{A}_i\}_{i\in[\tau]}\) are mutually independent, and each has bias at most \(1-2^{-d}\) since it is a non-constant deterministic function of \(d\) uniform bits. Thus no matter how \(\mathbf{B},\overline{\mathbf{Z}}\) are fixed, \(\X\) becomes a \((\delta,t)\)-biased \(d\)-local NOBF source, for \(\delta\leq1-2^{-d}\).
\end{proof}

\subsubsection{A reduction from biased \(d\)-local NOBF sources to (unbiased) \(d\)-local NOBF sources}\label{subsubsec:second-part-reduction}

We now prove the second part of our reduction (\cref{lem:simulate-coins}), which shows that every biased \(d\)-local NOBF source is (close to) a convex combination of (unbiased) \(d\)-local NOBF sources.

\begin{proof}[Proof of \cref{lem:simulate-coins}]
Let \(\X\sim\zo^n\) be a \((\delta,k)\)-biased \(d\)-local NOBF source. We wish to show that \(\X\) is \(\eps\)-close to a convex combination of \((0,k^\pr)\)-biased \(d\)-local NOBF sources, where \(k^\pr\geq(1-\delta)k/4\) and \(\eps=2^{-(1-\delta)k/4}\).

Without loss of generality, assume that the first \(k\) bits in \(\X\) are the ``good bits'': that is, there exist \(d\)-local functions \(g_{k+1},\dots,g_n : \zo^k\to\zo\) such that
\[
\X=(\X_1,\dots,\X_k,g_{k+1}(\X_1,\dots,\X_k),\dots,g_n(\X_1,\dots,\X_k),
\]
where each \(\X_i\) is independent and has bias at most \(\delta\). To remove the bias from this source, the key idea will be to simulate each \(\X_i\) by two independent coins: one which is biased, and one which is not.

In more detail, for every \(i\in[k]\) we construct a pair of independent random variables \(\mathbf{B}_i,\mathbf{A}_i\sim\zo\) as follows. First, let \(\gamma_i\in\zo\) be the value favored by \(\X_i\), breaking ties arbitrarily. Then, define \(p_i:=\Pr[\X_i=\gamma_i]\), and notice that \(\frac{1}{2}\leq p_i\leq\frac{1+\delta}{2}\), where the lower bound holds because \(\X_i\) favors \(\gamma_i\) over \(1-\gamma_i\), and the upper bound holds because the good bits have bias at most \(\delta\). Next, define \(\B_i\) such that
\begin{align*}
    \Pr[\mathbf{B}_i=0]&=2p_i-1,\\
    \Pr[\mathbf{B}_i=1]&=2-2p_i.
\end{align*}
Finally, let \(\mathbf{A}_i\sim\zo\) simply be a uniform bit, and define the function \(h_i:\zo\times\zo\to\zo\) as
\[
h_i(b,a):=
\begin{cases} 
    \gamma_i &\textbf{ if }b=0,\\
     a &\textbf{ if }b=1.
   \end{cases}
\]

Given these definitions, it is straightforward to verify that \(\X_i\) has the same distribution as \(h_i(\mathbf{B}_i,\mathbf{A}_i)\), and thus we may rewrite \(\X\) as
\[
(h_1(\mathbf{B}_1,\mathbf{A}_1), \dots, h_k(\mathbf{B}_k,\mathbf{A}_k), g_{k+1}(h_1(\mathbf{B}_1,\mathbf{A}_1), \dots, h_k(\mathbf{B}_k,\mathbf{A}_k)), \dots, g_n(h_1(\mathbf{B}_1,\mathbf{A}_1), \dots, h_k(\mathbf{B}_k,\mathbf{A}_k))).
\]

Now, define the random variable \(\mathbf{B}=(\mathbf{B}_1,\dots,\mathbf{B}_k)\). Given the above description of \(\X\), it is not too difficult to see that for any \(b\in\zo^k\), the conditional distribution \((\X\mid\mathbf{B}=b)\) is a \(d\)-local (unbiased) NOBF source, which has min-entropy equal to the Hamming weight of \(b\). Thus, we may write \(\X\) as the convex combination
\[
\X=\sum_{b\in\zo^k}\Pr[\mathbf{B}=b]\cdot(\X\mid\mathbf{B}=b).
\]
This means that if \(\mathbf{B}\) has Hamming weight \(\geq k^\pr\) with probability \(\geq1-\eps\), then \(\mathbf{X}\) is \(\eps\)-close to a convex combination of \(d\)-local NOBF sources with min-entropy \(\geq k^\pr\). Such a claim will follow almost immediately from a Chernoff bound.

In more detail, define a random variable \(\mathbf{Z}:=\sum_{i\in[k]}\mathbf{B}_i\) and notice that the value of \(\mathbf{Z}\) is exactly the Hamming weight of \(\mathbf{B}\). Furthermore, recall that \(\mathbf{B}_i=1\) with probability \(2-2p_i\) for some \(p_i\in[\frac{1}{2},\frac{1+\delta}{2}]\). Thus
\[
\mu:=\E[\mathbf{Z}]=\sum_{i\in[k]}\E[\mathbf{B}_i]\geq k\cdot(2-2\cdot((1+\delta)/2))=(1-\delta)k.
\]
Thus, by a standard Chernoff bound on the lower tail (\cref{lem:chernoff}), we get that
\[
\Pr[\mathbf{Z}\leq \mu/4]\leq e^{-\mu\cdot(3/4)^2/2}\leq 2^{-\mu/4},
\]
which means that \(\mathbf{B}\) has Hamming weight \(\geq\mu/4\) with probability \(\geq1-2^{-\mu/4}\). Thus, \(\X\) is \(2^{-\mu/4}\)-close to a convex combination of \(d\)-local NOBF sources with min-entropy \(\geq\mu/4\), as desired.
\end{proof}

\subsection{Low-degree polynomials extract from \(d\)-local NOBF sources}\label{subsec:upper-bounds:second-key-ingredient}

Now that we have proven our reduction from \(d\)-local sources to \(d\)-local NOBF sources (\cref{thm:upper-bounds:first-key-ingredient}), all that remains is to prove an upper bound on the entropy required to extract from \(d\)-local NOBF sources using degree \(\leq r\) polynomials (\cref{thm:upper-bounds:second-key-ingredient}). In this section, we prove \cref{thm:upper-bounds:second-key-ingredient}.

In order to prove \cref{thm:upper-bounds:second-key-ingredient}, our main tool will be the following lemma, which shows that for \emph{any} fixed NOBF source, a random low-degree polynomial can extract from it with very high probability.

\begin{lemma}\label{lem:random-extractor}
There is a universal constant \(c>0\) such that for any \(r\leq ck^{1/4}\), the following holds. Let \(\X\sim\zo^n\) be an NOBF source with min-entropy \(k\), and let \(f:\zo^n\to\zo\) be a random \(\F_2\)-polynomial of degree \(\leq r\). Then
\[
\Pr_f[f\text{ is an extractor for \(\X\) with error \(\eps=2^{-ck/r^3}\)}]\geq1-2^{-c{k\choose\leq r}}
\]
\end{lemma}

Before we prove this lemma, we show how it can be used to prove \cref{thm:upper-bounds:second-key-ingredient}. The proof is a straightforward application of the probabilistic method, which combines \cref{lem:random-extractor} with a basic upper bound on the number of \(d\)-local NOBF sources over \(\zo^n\) with min-entropy \(k\).

\begin{proof}[Proof of \cref{thm:upper-bounds:second-key-ingredient}]
Let \(\mathcal{X}\) be the family of \(d\)-local NOBF sources over \(\zo^n\) with min-entropy \(k\). Notice
\[
|\mathcal{X}|\leq{n\choose k}\cdot\left({k\choose d}\cdot2^{2^d}\right)^{n-k}.
\]
By \cref{lem:random-extractor}, a random degree \(\leq r\) polynomial \(f : \zo^n\to\zo\) fails to extracts from any fixed NOBF source of min-entropy \(k\) with error \(\eps=2^{-ck/r^3}\) with probability at most \(2^{-c{k\choose\leq r}}\). Thus, a union bound shows that a random degree \(\leq r\) polynomial \(f\) extracts from all sources in \(\mathcal{X}\) with error \(\eps=2^{-ck/r^3}\) with probability at least \(0.99\) (over the selection of \(f\)) as long as
\[
|\mathcal{X}|\cdot2^{-c{k\choose\leq r}}\leq{n\choose k}\cdot\left({k\choose d}\cdot2^{2^d}\right)^{n-k}\cdot2^{-c{k\choose\leq r}}\leq0.01,
\]
which is true if
\[
k\geq Cr \cdot (2^d\cdot n\log n)^{1/r}
\]
for a large enough constant \(C\).
\end{proof}

Now, all that remains is to prove \cref{lem:random-extractor}, which says that for any fixed NOBF source, a random low-degree polynomial can extract from it. In order to prove this result, our main tool will be following, which shows that for any fixed function \(g\), a random low-degree polynomial \(f\) has very low correlation with it.

\begin{lemma}\label{lem:main-correlation-bound}
There is a universal constant \(c>0\) such that for any fixed \(g:\F_2^n\to\F_2\), the following holds. For any \(1\leq r\leq cn^{1/4}\), let \(f:\F_2^n\to\F_2\) be a random degree \(\leq r\) polynomial. Then
\[
\Pr[\mathsf{corr}(f,g)>2^{-cn/r^3}]\leq2^{-c{n\choose\leq r}}.
\]
\end{lemma}

Before we prove these correlation bounds (\cref{lem:main-correlation-bound}), we show how they can be used to show that for any fixed NOBF source, a random low-degree polynomial can extract from it (\cref{lem:random-extractor}). Then, we conclude the section in \cref{subsubsec:correlation-bounds-sec} by proving \cref{lem:main-correlation-bound}.

\begin{proof}[Proof of \cref{lem:random-extractor}]
The proof is via the probabilistic method over the selection of \(f\). In more detail, let \(\X\sim\zo^n\) be an arbitrary NOBF source with min-entropy \(k\), and let \(f:\zo^n\to\zo\) be a random \(\F_2\)-polynomial of degree \(\leq r\). Assume (without loss of generality) that \(\X_1,\dots,\X_k\) are the ``good bits'' in the NOBF source.

Now, let \(f^\pr\) be the (sum of the) monomials in \(f\) that do not use any variables outside \(\X_1,\dots,\X_k\), and let \(g\) be the (sum of the) monomials in \(f\) that use at least one variable outside \(\X_1,\dots,\X_k\). This gives us \(f(\X)=f^\pr(\X)+g(\X)\), where \(f^\pr\) is a random degree \(\leq r\) polynomial over \(\X_1,\dots,\X_k\), and \(g\) is an (independent) random polynomial over \(\X_1,\dots,\X_n\). By definition of NOBF source, each of the bits \(\X_{k+1},\dots,\X_n\) is a deterministic function of the good bits \(\X_1,\dots,\X_k\), and thus we have
\[
f(\X)=f^\pr(\X_1,\dots,\X_k)+g^\pr(\X_1,\dots,\X_k),
\]
where \(f^\pr\) is a random degree \(\leq r\) polynomial over \(\X_1,\dots,\X_k\) and \(g^\pr\) is an independent (but not necessarily uniform) random polynomial over the same set of variables.

Since \(\X_1,\dots,\X_k\) are independent uniform bits, we get that \(f\) is an extractor with error
\begin{align*}
    \eps=\left|\Pr_{x\sim\U_k}[f(x)=1]-\frac{1}{2}\right|=\left|\Pr_{x\sim\U_k}[f^\pr(x)\neq g^\pr(x)]-\frac{1}{2}\right|=\frac{1}{2}\corr(f^\pr,g^\pr).
\end{align*}
Thus
\begin{align*}
	\Pr_f[f\text{ is \emph{not} an extractor for \(\X\) with error \(\eps\)}]=\Pr_{f^\pr,g^\pr}[\corr(f^\pr,g^\pr)>2\eps]\leq\Pr_{f^\pr}[\corr(f^\pr,g^\ast)>2\eps],
\end{align*}
where \(g^\ast=\argmax_{g^\pr}\Pr_{f^\pr}[\corr(f^\pr,g^\pr)>2\eps]\). Thus \(g^\ast:\F_2^k\to\F_2\) is an arbitrary fixed function and \(f^\pr:\F_2^k\to\F_2\) is a random degree \(\leq r\) polynomial. Thus, by combining the above inequality with \cref{lem:main-correlation-bound},
\begin{align*}
\Pr_f[f\text{ is \emph{not} an extractor for \(\X\) with error \(2^{-ck/r^3}\)}]&\leq \Pr_{f^\pr}[\corr(f^\pr,g^\ast)>2\cdot2^{-ck/r^3}]\\
&\leq\Pr_{f^\pr}[\corr(f^\pr,g^\ast)>2^{-ck/r^3}]\\
&\leq2^{-c\binom{k}{\leq r}},
\end{align*}
which completes the proof.
\end{proof}

\subsubsection{Correlation bounds against a single arbitrary function}\label{subsubsec:correlation-bounds-sec}

At last, all that remains in our proof of \cref{thm:upper-bounds:second-key-ingredient} is a proof of \cref{lem:main-correlation-bound}, which says that for any fixed function \(g\), a random low-degree polynomial \(f\) has very low correlation with it. As it turns out, \cref{lem:main-correlation-bound} follows quite readily from the following bounds on the list-size of Reed-Muller codes.

\begin{theorem}[\hspace{1sp}\cite{KLP12}]\label{thm:list-decodability-Reed-Muller}
	There is a universal constant \(C>0\) such that for any \(n,r\in\N\) and \(\eps>0\),	the Reed-Muller code \(\mathsf{RM}(n,r)\) is \((\frac{1-\eps}{2},L)\)-list-decodable, where
	\[
	\log L\leq C\cdot r\cdot(r+\log(1/\eps))\cdot{n\choose\leq r-1}.
	\]
\end{theorem}

Equipped with these bounds, we proceed to prove \cref{lem:main-correlation-bound}.

\begin{proof}[Proof of \cref{lem:main-correlation-bound}]
Let \(N:=2^n\), and consider any arbitrary functions \(f,g:\zo^n\to\zo\). If we let \(\widehat{f},\widehat{g}\in\zo^N\) denote the truth tables of \(f,g\), notice that
\[
N\cdot\corr(f,g)=|2\Delta(\widehat{f},\widehat{g})-N|,
\]
which implies that \(\corr(f,g)\leq\eps\) if and only if \(\Delta(\widehat{f},\widehat{g})\in [\frac{1-\eps}{2}\cdot N,\frac{1+\eps}{2}\cdot N]\). Thus, if we let \(Q\subseteq\zo^N\) denote the Reed-Muller code \(\mathsf{RM}(n,r)\), the following holds: for any fixed function \(g\) and a random degree \(\leq r\) polynomial \(f\),
\begin{align*}
\Pr_f[\corr(f,g)>\eps]&=\Pr_{q\sim Q}\left[\Delta(q,\widehat{g})\notin\left[\frac{1-\eps}{2}N, \frac{1+\eps}{2}N\right]\right]\\
&\leq\max_{v\in\zo^N}\Pr_{q\sim Q}\left[\Delta(q,v)\notin\left[\frac{1-\eps}{2}N, \frac{1+\eps}{2}N\right]\right]
\end{align*}

Thus, we would like to upper bound the probability that a random Reed-Muller codeword \(q\in\zo^N\) is \emph{not} within relative Hamming distance \([(1-\eps)/2,(1+\eps)/2]\) from an arbitrary point \(v\in\zo^N\). Towards this end, suppose that the Reed-Muller code \(Q\) is \((\frac{1-\eps}{2},L)\)-list decodable. We claim that for any \(v\in\zo^N\),
\[
\Pr_{q\sim Q}\left[\Delta(q,v)\notin\left[\frac{1-\eps}{2}N, \frac{1+\eps}{2}N\right]\right]\leq\frac{2L}{|Q|}.
\]
Indeed, this holds for \emph{any} \((\frac{1-\eps}{2},L)\)-list decodable code: to see why, simply note that \(\Delta(q,v)\notin\left[\frac{1-\eps}{2}N,\frac{1+\eps}{2}N\right]\) if and only if \(q\) is in one of the (open) balls \(\mathcal{B}(v,\frac{1-\eps}{2}N)\) or \(\mathcal{B}(\overline{v},\frac{1-\eps}{2}N)\), where \(\overline{v}\in\zo^N\) denotes the vector \(v\) with all of its bits flipped. By definition of list-decodability, each of these balls contains \(\leq L\) codewords, and the claim follows.

To conclude, recall that the Reed-Muller code \(\mathsf{RM}(n,r)\) has size \(|Q|=2^{\binom{n}{\leq r}}\), and \cref{thm:list-decodability-Reed-Muller} tells us that, for any \(\eps>0\), \(Q\) is \((\frac{1-\eps}{2},L)\)-list decodable, where \(
\log L\leq C\cdot r\cdot(r+\log(1/\eps))\cdot\binom{n}{\leq r-1}
\) and \(C\) is a universal constant. It is a straightforward calculation to verify that there is a universal constant \(c>0\) such that for \(\eps=2^{-cn/r^3}\), it holds that
\[
\log L\leq C\cdot r\cdot (r+\log(1/\eps))\cdot\binom{n}{\leq r-1}\leq\binom{n}{\leq r}/2,
\]
as long as \(r\leq cn^{1/4}\). Combining all of our inequalities, we get that as long as \(r\leq cn^{1/4}\), it holds that for any fixed function \(g\) and a random degree \(\leq r\) polynomial \(f\),
\[
\Pr_f[\corr(f,g)>\eps=2^{-cn/r^3}]\leq\frac{2L}{|Q|}\leq\frac{2\cdot 2^{\binom{n}{\leq r}/2}}{2^{\binom{n}{\leq r}}}=2^{-\binom{n}{\leq r}/2+1}\leq 2^{-\binom{n}{\leq r}/4}\leq 2^{-c\binom{n}{\leq r}},
\]
as desired.
\end{proof}

\subsection{Optimized upper bounds for dispersers}\label{subsec:optimized-dispersers}

In this subsection, we show that we can improve our entropy upper bounds if we only wish to \emph{disperse} from \(d\)-local sources using degree \(\leq r\) polynomials (instead of \emph{extract}). We prove the following theorem, which is an optimized version of \cref{thm:localextintro:restated} for the setting of dispersers.

\begin{theorem}[Formal version of \cref{rem:main-remark-here}]\label{thm:formal-version-of-main-remark}
There is a universal constant \(C>0\) such that for all \(n,d,r\in\N\), the following holds. With probability at least \(0.99\) over the choice of a random degree \(\leq r\) polynomial \(f\in\F_2[x_1,\dots,x_n]\), it holds that \(f\) is a disperser for \(d\)-local sources with min-entropy \(\geq k\), as long as
\[
k\geq Cd^2r\cdot(dn\log n)^{1/r} + Cd^3\cdot n^{1/r}.
\]
\end{theorem}

As \cref{thm:formal-version-of-main-remark} is an optimized version of \cref{thm:localextintro:restated} for the setting of dispersers, we set out to prove \cref{thm:formal-version-of-main-remark} by re-examining our proof of \cref{thm:localextintro:restated}. Recall that \cref{thm:localextintro:restated} shows that low-degree polynomials can extract from \(d\)-local sources, and in order to prove it, we combined two main ingredients:
\begin{enumerate}
    \item A reduction from \(d\)-local sources to \(d\)-local NOBF sources (\cref{thm:upper-bounds:first-key-ingredient}).
    \item A result showing that low-degree polynomials can extract from \(d\)-local NOBF sources (\cref{thm:upper-bounds:second-key-ingredient}).
\end{enumerate}
In order to optimize \cref{thm:localextintro:restated} for the setting of dispersers (and thereby prove \cref{thm:formal-version-of-main-remark}), we show that both of these ingredients can be optimized for dispersers.

In more detail, in order to prove \cref{thm:formal-version-of-main-remark}, we combine two key ingredients. Just like the first key ingredient in \cref{thm:localextintro:restated} (\cref{thm:upper-bounds:first-key-ingredient}), our first key ingredient for \cref{thm:formal-version-of-main-remark} reduces \(d\)-local sources to the more specialized family of \(d\)-local NOBF sources. However, as we are dealing with the easier setting of \emph{dispersing} (instead of \emph{extracting}), the definition of \emph{reduction} here is not as strict.

\begin{theorem}[Optimized version of \cref{thm:upper-bounds:first-key-ingredient} for dispersers]\label{thm:upper-bounds:first-key-ingredient:optimized-for-dispersers}
There exists a universal constant \(c>0\) such that for any \(n,k,d\in\N\), the following holds. Let \(\X\sim\zo^n\) be a \(d\)-local source with min-entropy \(\geq k\). Then there is a \(d\)-local NOBF source \(\X^\pr\sim\zo^n\) with min-entropy \(\geq k^\pr\) such that \(\supp(\X^\pr)\subseteq\supp(\X)\), where
\[
k^\pr=\frac{ck}{d^2}.
\]
\end{theorem}

Next, while the second key ingredient in \cref{thm:localextintro:restated} (\cref{thm:upper-bounds:second-key-ingredient}) shows that a random low-degree polynomial \emph{extracts} from \(d\)-local NOBF sources, the second key ingredient for \cref{thm:formal-version-of-main-remark} just needs to show that a random low-degree polynomial \emph{disperses} from \(d\)-local NOBF sources. Given this weaker requirement, we are able to improve the min-entropy requirement and prove the following, which is our second key ingredient.

\begin{theorem}[Optimized version of \cref{thm:upper-bounds:second-key-ingredient} for dispersers]\label{thm:upper-bounds:second-key-ingredient:optimized-for-dispersers}
There is a universal constants \(C>0\) such that for all \(n,d,r\in\N\), the following holds.
With probability at least \(0.99\) over the choice of a random degree \(\leq r\) polynomial \(f\in\F_2[x_1,\dots,x_n]\), it holds that \(f\) is a disperser for \(d\)-local NOBF sources of min-entropy \(\geq k\), as long as
\[
k\geq Cr\cdot(d n\log n)^{1/r}+Cd\cdot n^{1/r}
\]
\end{theorem}

Now, recall that if a disperser works for a family \(\mathcal{X}\) of sources over \(\zo^n\), then it also works for any \(\X\sim\zo^n\) whose support contains the support of some \(\X^\pr\in\mathcal{X}\) (\cref{fact:convex-combo-lifted-disperser-optimized}). As a result, by combining \cref{thm:upper-bounds:first-key-ingredient:optimized-for-dispersers,thm:upper-bounds:second-key-ingredient:optimized-for-dispersers}, we immediately get \cref{thm:formal-version-of-main-remark}. Thus, in the remainder of this section, we just focus on proving \cref{thm:upper-bounds:first-key-ingredient:optimized-for-dispersers,thm:upper-bounds:second-key-ingredient:optimized-for-dispersers}. We prove \cref{thm:upper-bounds:first-key-ingredient:optimized-for-dispersers} in \cref{subsubsec:first-disperser-optimization}, and we prove \cref{thm:upper-bounds:second-key-ingredient:optimized-for-dispersers} in \cref{subsubsec:second-disperser-optimization}. This will conclude our discussion of entropy upper bounds (and \cref{sec:new-upper-bounds}).

\subsubsection{A reduction from \(d\)-local sources to \(d\)-local NOBF sources (optimized for dispersers)}\label{subsubsec:first-disperser-optimization}

We start by proving \cref{thm:upper-bounds:first-key-ingredient:optimized-for-dispersers}, a reduction from \(d\)-local sources to \(d\)-local NOBF sources, which is optimized for dispersers. As it turns out, \cref{thm:upper-bounds:first-key-ingredient:optimized-for-dispersers} follows quite readily from our reduction from \(d\)-local sources to biased \(d\)-local NOBF sources (\cref{lem:hard-part-of-reduction}).

\begin{proof}[Proof of \cref{thm:upper-bounds:first-key-ingredient:optimized-for-dispersers}]
Let \(\X\sim\zo^n\) be a \(d\)-local source with min-entropy \(k\). By \cref{lem:hard-part-of-reduction}, \(\X\) is a convex combination of \((\delta,k^\pr)\)-biased \(d\)-local NOBF souces, where \(\delta<1\) and \(k^\pr\geq k/(2d^2)\). Let \(\Y\sim\zo^n\) be an element of this convex combination (that is assigned probability \(>0\)). Clearly \(\Y\) is a \((\delta,k^\pr)\)-biased \(d\)-local NOBF source, and furthermore \(\supp(\Y)\subseteq\supp(\X)\).

Now, let \(\X^\pr\sim\zo^n\) be the ``unbiased'' version of \(\Y\): that is, let \(\X^\pr\) have the same set of \(k^\pr\) ``good'' bits as \(\Y\), and let the bad bits in \(\X^\pr\) depend on the good bits via the same deterministic functions as in \(\Y\). However, make the good bits in \(\X^\pr\) have \(0\) bias. Then \(\X^\pr\) is an (unbiased) \(d\)-local NOBF source with min-entropy \(k^\pr\), and \(\supp(\X^\pr)=\supp(\Y)\subseteq\supp(\X)\).
\end{proof}

\subsubsection{Low-degree polynomials disperse from \(d\)-local NOBF sources (optimized for dispersers)}\label{subsubsec:second-disperser-optimization}

Now that we have proven our reduction from \(d\)-local sources to \(d\)-local NOBF sources (\cref{thm:upper-bounds:first-key-ingredient:optimized-for-dispersers}), all that remains is to prove an upper bound on the entropy required to disperse from \(d\)-local NOBF sources using degree \(\leq r\) polynomials (\cref{thm:upper-bounds:second-key-ingredient:optimized-for-dispersers}). In this section, we prove \cref{thm:upper-bounds:second-key-ingredient:optimized-for-dispersers}.

In order to prove \cref{thm:upper-bounds:second-key-ingredient:optimized-for-dispersers}, our key ingredient is a lemma about \emph{function composition} and how it affects \emph{which degrees are ``hit.''} More formally, for any nonnegative integer \(r\), we say that a function \(f:\F_2^n\to\F_2\) \emph{hits degree \(r\)} if the unique multilinear \(\F_2\)-polynomial computing \(f\) includes a monomial of size \(r\),\footnote{The monomial of size \(0\) is the constant \(1\).} and we say that \(f\) \emph{hits degree \(\leq r\)} if there is some \(0\leq r^\pr\leq r\) such that \(f\) hits degree \(r^\pr\). Then, using standard notation for function composition, we prove the following.

\begin{lemma}\label{lem:hitting-lemma}
Let \(f:\F_2^n\to\F_2\) and \(a_1,\dots,a_n:\F_2^k\to\F_2\) be functions such that \(f(a_1,\dots,a_n)\) hits degree \(r\). Furthermore, let \(b_1,\dots,b_n:\F_2^k\to\F_2\) be functions that do not hit degree \(\leq r\). Then the function \(f(a_1+b_1,\dots,a_n+b_n):\F_2^k\to\F_2\) hits degree \(r\).
\end{lemma}

Before we prove this lemma, we show how it can be used to to show that degree \(\leq r\) polynomials disperse from \(d\)-local NOBF sources, proving \cref{thm:upper-bounds:second-key-ingredient:optimized-for-dispersers}. Then, we conclude this section by proving \cref{lem:hitting-lemma}.

\begin{proof}[Proof of \cref{thm:upper-bounds:second-key-ingredient:optimized-for-dispersers}]

We start with some basic definitions. First, recall that a \(d\)-local NOBF source of min-entropy \(k\) is a random variable \(\X\sim\zo^n\) that is generated as follows: there exist \(a_1,\dots,a_n:\F_2^k\to\F_2\) such that \(\X=(a_1(\U_k),\dots,a_n(\U_k))\), where each \(\U_k\) is a copy of the same uniform random variable. Furthermore, the following must hold: there exists a ``good'' set \(S=\{\alpha_1,\dots,a_k\}\subseteq[n]\) such that each \(a_{\alpha_i}\) simply outputs its \(i^\text{th}\) input, and for every other (\emph{bad}) coordinate \(j\notin S\) we have that \(a_j\) depends on just \(d\) inputs. For convenience, we refer to the set of functions \((a_1,\dots,a_n)\) as the \emph{generating functions} of the \(d\)-local NOBF source. If each \(a_i\) is a degree \(\leq r\) polynomial, we say that \(\X\) is not just a \(d\)-local NOBF source, but a \emph{(\(d\)-local, degree \(\leq r\)) NOBF source}.

Now, the goal is to use \cref{lem:hitting-lemma} to show that, in order for a function \(f:\F_2^n\to\F_2\) to disperse from all \(d\)-local NOBF sources of min-entropy \(k\), it suffices to show that \(f\) ``hits'' some degree in \([1,r]\) on all (\(d\)-local, degree \(\leq r\)) NOBF sources of the same min-entropy. Then, a straightforward application of the probabilistic method over the latter (smaller) family will yield the result.
\\

More formally, fix an arbitrary function \(f:\F_2^n\to\F_2\). Suppose that for any (\(d\)-local, degree \(\leq r\)) NOBF source \(\X\sim\zo^n\) with min-entropy \(k\) and generating functions \(a_1,\dots,a_n:\F_2^k\to\F_2\), there is some \(r^\pr\in[1,r]\) such that \(f(a_1,\dots,a_n)\) hits degree \(r^\pr\). Then, we \textbf{claim} that \(f\) is not only a disperser for all (\(d\)-local, degree \(\leq r\)) NOBF sources of min-entropy \(k\), but it is also a disperser for all \(d\)-local NOBF sources of min-entropy \(k\).

To see why, let \(\X^\pr\sim\zo^n\) be a \(d\)-local NOBF source with generating functions \(h_1,\dots,h_n:\F_2^k\to\F_2\), and for each \(i\in[n]\) define the functions \(a_i^\pr,b_i^\pr:\F_2^k\to\F_2\) such that \(a_i^\pr\) is the sum of all monomials in \(h_i\) of degree \(\leq r\), and \(b_i^\pr\) is the sum of all monomials in \(h_i\) of degree \(>r\). Consider now the source
\[
\X:=(a_1^\pr(\U_k),\dots,a_n^\pr(\U_k)),
\]
and notice that \(\X\) is a (\(d\)-local, degree \(\leq r\)) NOBF source with min-entropy \(k\) and generating functions \(a_1^\pr,\dots,a_n^\pr:\F_2^k\to\F_2\). By the hypothesis, \(f(a_1^\pr,\dots,a_n^\pr)\) hits degree \(r^\pr\), where \(r^\pr\in[1,r]\). And since \(b_1^\pr,\dots,b_n^\pr:\F_2^k\to\F_2\) do \emph{not} hit degree \(\leq r\), we know by \cref{lem:hitting-lemma} that \(f(a_1^\pr+b_1^\pr,\dots,a_n^\pr+b_n^\pr)=f(h_1,\dots,h_n)\) hits degree \(r^\pr\in[1,r]\). This means that \(f(h_1,\dots,h_n)\) is not constant, and since \(f(\X^\pr)=f(h_1,\dots,h_n)(\U_k)\), it holds that \(f(\X^\pr)\) is not constant: in other words, \(f\) is a disperser for \(\X^\pr\). Thus, \(f\) is a disperser for all \(d\)-local NOBF sources of min-entropy \(k\).
\\

Thus, all that remains is to show that with probability \(\geq0.99\) over the choice of a random degree \(\leq r\) polynomial \(f\in\F_2[x_1,\dots,x_n]\), the following holds: for any (\(d\)-local, degree \(\leq r\)) NOBF source \(\X\sim\zo^n\) with min-entropy \(k\) and generating functions \(a_1,\dots,a_n:\F_2^k\to\F_2\), there is some \(r^\pr\in[1,r]\) such that \(f(a_1,\dots,a_n)\) hits degree \(r^\pr\).
\\

Towards this end, let \(f\in\F_2[x_1,\dots,x_n]\) be a (uniformly) random degree \(\leq r\) polynomial, and let \(\X\sim\zo^n\) be an arbitrary NOBF source with min-entropy \(k\) and generating functions \(a_1,\dots,a_n:\F_2^k\to\F_2\). Without loss of generality, assume that \(a_i(y)=y_i\) for all \(i\in[k]\). Now, we want to upper bound the probability that for all \(r^\pr\in[1,r]\), it holds that \(f(a_1,\dots,a_n)\) does \emph{not} hit degree \(r^\pr\). We \textbf{claim} that this probability is at most \(2^{-\binom{k}{\leq r}+1}\).

To see why, note that since \(f:\F_2^n\to\F_2\) is a uniformly random degree \(\leq r\) polynomial, we have
\[
f(x):=\sum_{S\subseteq[n] : |S|\leq r}c_Sx^S,
\]
where each \(c_S\sim\zo\) is independent and uniform. Now, note that for \(y\in\F_2^k\),
\begin{align*}
    f(a_1,\dots,a_n)(y)&=\sum_{S\subseteq[n]:|S|\leq r}c_S a^S(y)\\
    &=\sum_{S\subseteq[k]:|S|\leq r}c_Sa^S(y) + \sum_{S\subseteq[n]:|S|\leq r,S\not\subseteq[k]}c_Sa^S(y)\\
    &=\sum_{S\subseteq[k]:|S|\leq r}c_Sy^S+\sum_{S\subseteq[n]:|S|\leq r, S\not\subseteq[k]}c_Sa^S(y).
\end{align*}
Now, notice that for each fixing of \(\{c_S\}_{S\subseteq[n]:|S|\leq r,S\not\subseteq[k]}\), we get that
\[
f(a_1,\dots,a_n)(y)=f^\pr(y)+g(y),
\]
where \(f^\pr:\F_2^k\to\F_2\) is a uniformly random polynomial of degree at most \(\leq r\), and \(g:\F_2^k\to\F_2\) is a fixed polynomial. Now, for any \(S\subseteq[k]\) of size at most \(r\), the probability that \(f^\pr+g\) \emph{does not} include the monomial \(y^S\) is the probability that \(f^\pr\) copies the decision of \(g\) on whether or not to include \(y^S\): this probability is \(1/2\). More generally, the probability that \(f^\pr+g\) \emph{does not} hit degree \(r\) will be \(2^{-{k\choose r}}\), and the probability that it does not hit any degree in \([1,r]\) will be \(2^{-{k\choose\leq r}+1}\). Thus
\begin{align}\label{eq:single-hit-failure}
\Pr_f[f(a_1,\dots,a_n)(x)\text{ does not hit any degree in }[1,r]]\leq2^{-{k\choose\leq r}+1},
\end{align}
as claimed.
\\

Now, we say that \(f\) \emph{fails} if there exists a (\(d\)-local, degree \(\leq r\)) NOBF source \(\X\sim\zo^n\) of min-entropy \(k\) with generating functions \(a_1,\dots,a_n\) such that \(f(a_1,\dots,a_n)\) does not hit any degree in \([1,r]\). By combining \cref{eq:single-hit-failure} with a simple union bound, we get that
\[
\Pr_f[f\text{ fails}]\leq|\mathcal{X}|\cdot2^{-\binom{k}{\leq r}+1},
\]
where \(\mathcal{X}\) is the family of (\(d\)-local, degree \(\leq r\)) NOBF sources \(\X\sim\zo^n\) of min-entropy \(k\). To upper bound the size of \(\mathcal{X}\), simply note that each \(\X\in\mathcal{X}\) can: (i) choose the locations of its \(k\) good bits; (ii) choose the \(d\) good bits that each bad bit depends on; and (iii) choose the degree \(\leq r\) polynomial that computes each bad bit. Thus
\[
|\mathcal{X}|\leq \binom{n}{k}\cdot\left(\binom{k}{d}\cdot2^{\binom{d}{\leq r}}\right)^{n-k}.
\]
It is now a straightforward calculation to verify that there is a sufficiently large constant \(C>0\) such that for all \(k\geq Cr(d n\log n)^{1/r}+Cdn^{1/r}\), it holds that
\[
\Pr_f[f\text{ fails}]\leq|\mathcal{X}|\cdot2^{-\binom{k}{\leq r}+1}\leq \binom{n}{k}\cdot\left(\binom{k}{d}\cdot2^{\binom{d}{\leq r}}\right)^{n-k}\cdot2^{-\binom{k}{\leq r}+1}\leq0.01.
\]

To conclude, we get that with probability \(\geq0.99\) over the choice of a random degree \(\leq r\) polynomial \(f\in\F_2[x_1,\dots,x_n]\), it holds that \(f\) hits some degree in \([1,r]\) on every (\(d\)-local, degree \(\leq r\)) NOBF source with min-entropy \(k\), provided that \(k\geq Cr\cdot(dn\log n)^{1/r}\). By our claim from the beginning of the proof, it follows that \(f\) is also a disperser for all \(d\)-local NOBF sources of min-entropy \(k\), as desired.
\end{proof}

At last, all that remains is to prove \cref{lem:hitting-lemma}, which says that if \(f:\F_2^n\to\F_2\) and \(a_1,\dots,a_n:\F_2^k\to\F_2\) are functions such that \(f(a_1,\dots,a_n)\) hits degree \(r\), and \(b_1,\dots,b_n:\F_2^k\to\F_2\) are functions that do not hit degree \(\leq r\), then \(f(a_1+b_1,\dots,a_n+b_n)\) hits degree \(r\).

\begin{proof}[Proof of \cref{lem:hitting-lemma}]
Since we can write any function \(f:\F_2^n\to\F_2\) as a unique multilinear  \(\F_2\)-polynomial, we know that
\[
f(x)=\sum_{S\subseteq[n]}c_S x^S
\]
for some constants \(c_S\in\zo\), \(S\subseteq[n]\). Using this, we have
\begin{align*}
f(a_1+b_1,\dots,a_n+b_n)(x)&=\sum_{S\subseteq[n]}c_S\prod_{i\in S}(a_i(x)+b_i(x))\\
&=\sum_{S\subseteq[n]}c_S\sum_{T\subseteq S}a^T(x)b^{S-T}(x)\\
&=\sum_{S\subseteq[n]}c_S\left(a^S(x)+\sum_{T\subsetneq S}a^T(x)b^{S-T}(x)\right)\\
&=\sum_{S\subseteq[n]}c_Sa^S(x) +\sum_{S\subseteq[n]}\sum_{T\subsetneq S}a^T(x)b^{S-T}(x)\\
&=f(a_1,\dots,a_n)(x)+\sum_{S\subseteq[n]}\sum_{T\subsetneq S}a^T(x)b^{S-T}(x)\\
&=f(a_1,\dots,a_n)(x)+\sum_{\substack{\emptyset\subsetneq S\subseteq[n]\\T\subsetneq S}}a^T(x)b^{S-T}(x).
\end{align*}

Now, for each \(S\subseteq[n],T\subsetneq S\) such that \(S\) is nonempty, define the function \(g_{S,T}(x):=a^T(x)b^{S-T}(x)\). Since \(f(a_1,\dots,a_n)\) hits degree \(r\), notice that in order to complete the proof, it suffices to show that each \(g_{S,T}\) does \emph{not} hit degree \(r\). Towards this end, pick an arbitrary element \(\gamma\in S-T\), and observe
\begin{align*}
    g_{S,T}(x):&=a^T(x)b^{S-T}(x)\\
    &=\prod_{i\in T}a_i(x)\prod_{j\in S-T}b_j(x)\\
    &=\left(\prod_{i\in T}a_i(x)\prod_{j\in (S-T)-\{\gamma\}}b_j(x)\right)b_\gamma(x).
\end{align*}
Next, it is straightforward to verify that for any functions \(p,q:\F_2^k\to\F_2\) such that \(q\) does not hit degree \(\leq r\), it follows that the function \(pq:\F_2^k\to\F_2\) does not hit degree \(\leq r\).\footnote{This is because any monomial that ``pops out'' of \(pq\) must be at least the size of the smallest monomial in \(q\), which is \(>r\).} Thus, since \(b_\gamma\) does not hit degree \(\leq r\) by the hypothesis, it follows via the above observation and identity that \(g_{S,T}\) does not hit degree \(\leq r\), which is even stronger than what we needed to complete the proof.
\end{proof}

\dobib

\section{Entropy lower bounds}\label{sec:lower-bounds-new}

In this section, we obtain lower bounds on the entropy required to extract from \(d\)-local sources using degree \(\leq r\) polynomials, proving the following theorem.

\begin{theorem}[\cref{thm:localtightintro}, restated]\label{thm:localtightintro:restated}
There is a constant \(c>0\) such that for all \(n,d,r\in\N\) with \(2\leq r\leq c\log(n)\) and \(d\leq n^{\frac{1}{r-1}-2^{-10r}}/\log(n)\), the following holds. For any degree \(\leq r\) polynomial \(f\in\F_2[x_1,\dots,x_n]\), there is a \(d\)-local source \(\X\sim\zo^n\) with min-entropy at least
\[
k\geq cr (dn\log n)^{1/r}
\]
such that \(f(\X)\) is constant.
\end{theorem}

In order to prove \cref{thm:localtightintro:restated}, we actually prove a stronger theorem, which can be viewed as a \emph{local} version of a result by Cohen and Tal~\cite{CT15}. In particular, Cohen and Tal show that every \(n\)-variate polynomial of degree \(\leq r\) admits a subspace of dimension \(\Omega(rn^{\frac{1}{r-1}})\) on which it is constant, and that this is tight. We show that a similar result holds even for the special subclass of \emph{local} subspaces.

\begin{theorem}[\cref{thm:local-cohen-tal}, restated]\label{thm:local-cohen-tal:restated}
There exist universal constants \(C,c>0\) such that for every \(n,r,d\in\N\) such that \(2\leq r\leq c\log(n)\) and \(d\leq n^{\frac{1}{r-1}-2^{-10r}}/\log(n)\), the following holds. For any degree \(\leq r\) polynomial \(f\in\F_2[x_1,\dots,x_n]\), there exists a \(d\)-local subspace \(X\subseteq\F_2^n\) of dimension
\[
k\geq cr (d n\log n)^{1/r}
\]
on which \(f\) is constant.

This is tight: there exists a degree \(\leq r\) polynomial \(g\in\F_2[x_1,\dots,x_n]\) which is an extractor for \(d\)-local affine sources of dimension \(k\geq Cr(dn\log n)^{1/r}\), which has error \(\eps=2^{-ck/r}\).
\end{theorem}

Notice that \cref{thm:local-cohen-tal:restated} immediately implies \cref{thm:localtightintro:restated}, since (the uniform distribution over) a \(d\)-local subspace is not only an affine source, but it is also a \(d\)-local source. Thus, in the remainder of the section, we focus on proving \cref{thm:local-cohen-tal:restated}. The key new ingredient we rely on is a so-called ``low-weight Chevalley-Warning theorem,'' which may be of independent interest. We prove this result in \cref{subsec:low-wt-cw}, and we show how it can be used to obtain \cref{thm:local-cohen-tal:restated} in \cref{subsec:cohen-tal-local-sec}. This will conclude our discussion of entropy lower bounds (and \cref{sec:lower-bounds-new}).

\subsection{A low-weight Chevalley-Warning theorem}\label{subsec:low-wt-cw}

The \emph{Chevalley-Warning theorem} is a classical result from number theory, which guarantees that a small set of low-degree polynomials admits a nontrivial common solution. More formally, it states the following.

\begin{theorem}[Chevalley-Warning theorem~\cite{War36}]\label{thm:absolutely-standard-Chevalley-Warning}
Let \(\{f_i\}\subseteq\F_2[x_1,\dots,x_n]\) be a set of polynomials with degree at most \(D\) such that \(0\) is a common solution. Then there are at least \(2^{n-D}\) common solutions to \(\{f_i\}\). In particular, if \(D<n\), then there must be a nontrivial common solution.
\end{theorem}

In this subsection, we prove a ``low-weight'' version of this theorem (which will be instrumental in our proof of \cref{thm:local-cohen-tal:restated}). In more detail, we believe it is natural to ask not only if \(\{f_i\}\) contains a nontrivial common solution, but if \(\{f_i\}\) contains a nontrivial common solution of \emph{low Hamming weight}.

In the case where all the \(f_i\) are linear, this is a question about the distance-dimension tradeoff of linear codes. Here, this question is answered by the \emph{Hamming bound}, which says that there must be a nontrivial common solution of weight \(w\leq O(D/\log(n/D))\).\footnote{More formally, it says that if all nontrivial solutions have weight \(>w\), then it must hold that \(\binom{n}{\leq\lfloor w/2\rfloor}\leq 2^D\).}

On the other hand, for general collections \(\{f_i\}\) that may have nonlinear polynomials, it is straightforward to use classical Chevalley-Warning (\cref{thm:absolutely-standard-Chevalley-Warning}) to show that there is always a nontrivial common solution of weight \(w\leq D+1\),\footnote{Given a collection \(\{f_i\}\subseteq\F_2[x_1,\dots,x_n]\) of degree \(D\), apply \cref{thm:absolutely-standard-Chevalley-Warning} to the collection \(\{f_i\}\cup\{x_1,x_2,\dots,x_{n-(D+1)}\}\).} and one can show that this is tight in general.\footnote{Consider the singleton set \(\{f\}\subseteq\F_2[x_1,\dots,x_n]\), where \(f(x)=\sum_{\emptyset\subsetneq S\subseteq[n]:|S|\leq D}x^S\).}

Thus, the upper bound on Hamming weight is much better when we know that all the polynomials in \(\{f_i\}\) are linear. This begs the question: if we know that \emph{most} of the polynomials in \(\{f_i\}\) are linear, can we get an upper bound on Hamming weight that is almost as strong as in the purely linear case? And, more generally, can we get a granular bound that takes into account exactly how many \(f_i\) are linear and nonlinear?

Our \emph{low-weight Chevalley-Warning theorem}, which we state next, provides a result of exactly this form.

\begin{theorem}\label{thm:low-wt-cw:technical}
Let \(\{f_i\}\subseteq\F_2[x_1,\dots,x_n]\) be a set of polynomials with linear degree at most \(D\) and nonlinear degree at most \(\Delta\), such that \(0\) is a common solution and \(D+\Delta<n\). If all nontrivial common solutions to \(\{f_i\}\) have Hamming weight \(>w\), then
\begin{align*}
\binom{n}{\leq\lfloor w/2\rfloor}\leq2^{D+\Delta+1}\cdot\binom{n}{\leq\lfloor\Delta/2\rfloor}.
\end{align*}
\end{theorem}

As we will see, a relatively straightforward (but tedious) calculation yields the following corollary.

\begin{corollary}[\cref{thm:low-wt-cw}, restated]\label{cor:low-wt-cw:main-technical-cor}
Let \(\{f_i\}\subseteq\F_2[x_1,\dots,x_n]\) be a set of polynomials with linear degree at most \(D\) and nonlinear degree at most \(\Delta\) such that \(0\) is a common solution and \(D+\Delta<n\). Then there is a nontrivial common solution with Hamming weight
\[
w\leq 8\Delta + 8D/\log(n/D)+8.\footnote{We define the expression \(8D/\log(n/D)\) to be \(0\) if \(D=0\).}
\]
\end{corollary}

Notice that the weight upper bound in \cref{cor:low-wt-cw:main-technical-cor} tightly interpolates (up to constant factors) between the linear and nonlinear cases discussed at the beginning of this section. We will also see that it is not difficult to extend \cref{cor:low-wt-cw:main-technical-cor} to obtain the following, which is the result that we will actually end up using in our proof of \cref{thm:local-cohen-tal:restated}.

\begin{corollary}\label{cor:projection-cw}
Let \(\{f_i\}\subseteq\F_2[x_1,\dots,x_n]\) be a set of polynomials with linear degree at most \(D\) and nonlinear degree at most \(\Delta\) such that \(0\) is a common solution. Then for any \(S\subseteq[n]\) such that \(D+\Delta<|S|\), there is a nontrivial common solution supported on \(S\) with Hamming weight
\[
w\leq 8\Delta + 8D/\log(|S|/D) + 8.
\]
\end{corollary}

In the remainder of this section, we prove \cref{thm:low-wt-cw:technical} and \cref{cor:low-wt-cw:main-technical-cor,cor:projection-cw}. The proof of \cref{thm:low-wt-cw:technical} is most interesting, and can be found in \cref{subsubsec:proof-low-wt-cw:main}. The proofs of \cref{cor:low-wt-cw:main-technical-cor,cor:projection-cw} are less interesting and somewhat tedious, and can be found in \cref{subsubsec:proof-low-wt-cw:corollaries}.

\subsubsection{A proof of our low-weight Chevalley-Warning theorem}\label{subsubsec:proof-low-wt-cw:main}

We now prove our low-weight Chevalley-Warning theorem (\cref{thm:low-wt-cw:technical}). The key ingredient we need is the following lemma, which says that for any collection of polynomials \(\{f_i\}\) and any big enough set of common solutions \(A\), it holds that \(A+A\) contains a nontrivial common solution.

\begin{lemma}\label{lem:key-ingredient:low-wt-cw:technical}
Let \(\{f_i\}\subseteq\F_2[x_1,\dots,x_n]\) be a set of polynomials with nonlinear degree at most \(\Delta\) such that \(0\) is a common solution. Then for any set \(A\subseteq\F_2^n\) of common solutions of size
\[
|A|>2\binom{n}{\leq\lfloor\Delta/2\rfloor},
\]
it holds that \(A+A\) contains a nontrivial common solution.
\end{lemma}

Before proving this result, we show how it can be combined with the Hamming bound and the classical Chevalley-Warning theorem to get \cref{thm:low-wt-cw:technical}, our low-weight Chevalley-Warning theorem.

\begin{proof}[Proof of \cref{thm:low-wt-cw:technical}]
Let \(Q\subseteq\F_2^n\) be the set of common solutions to \(\{f_i\}\), and suppose that all \(v\in Q-\{0\}\) have Hamming weight \(>w\). Then for any ball \(\mathcal{B}\) of radius \(\lfloor w/2\rfloor\), it must hold that
\[
|Q\cap\mathcal{B}|\leq 2\binom{n}{\leq\lfloor\Delta/2\rfloor},
\]
since otherwise \cref{lem:key-ingredient:low-wt-cw:technical} (combined with the triangle inequality) implies that that there is a nontrivial solution of weight at most \(2\lfloor w/2\rfloor\leq w\). Thus, we see that \(Q\) is a \((\rho,L)\)-list-decodable code, where
\begin{align*}
    \rho &= \lfloor w/2\rfloor / n,\\
    L &= 2\binom{n}{\leq\lfloor\Delta/2\rfloor}.
\end{align*}
Now, the Hamming bound (\cref{thm:hamming-bound}) implies that \(|Q|\leq 2^nL/\binom{n}{\leq\rho n}\), whereas the Chevalley-Warning theorem (\cref{thm:absolutely-standard-Chevalley-Warning}) implies that \(2^{n-(D+\Delta)}\leq |Q|\). Combining these inequalities yields
\[
2^{n-(D+\Delta)}\leq \frac{2^n L}{\binom{n}{\leq\rho n}}\leq \frac{2^n\cdot 2\binom{n}{\leq\lfloor\Delta/2\rfloor}}{\binom{n}{\leq\lfloor w/2\rfloor}},
\]
which immediately implies the result.
\end{proof}

All that remains is to prove our key ingredient, \cref{lem:key-ingredient:low-wt-cw:technical}. As it turns out, it follows quite readily from the following \emph{CLP lemma}, which was instrumental in the recent resolution of the cap set conjecture.

\begin{lemma}[\hspace{1sp}\cite{CLP17}]\label{lem:clp}
Let \(f\in\F_2[x_1,\dots,x_n]\) be a polynomial of degree at most \(r\), and let \(M\) denote the \(2^n\times 2^n\) matrix with entries \(M_{x,y}=f(x+y)\) for \(x,y\in\F_2^n\). Then
\[
\mathsf{rank}(M)\leq2\binom{n}{\leq\lfloor r/2\rfloor}.
\]
\end{lemma}

Given the CLP lemma, we are ready to prove \cref{lem:key-ingredient:low-wt-cw:technical}, which will conclude our proof of \cref{thm:low-wt-cw:technical}.

\begin{proof}[Proof of \cref{lem:key-ingredient:low-wt-cw:technical}]
First, let \(\{g_i\}\subseteq\{f_i\}\) be the set of polynomials in \(\{f_i\}\) that have degree \(>1\). Notice that if \(A+A\) contains a nontrivial common solution to the system \(\{g_i\}\), then it also contains a nontrivial common solution to \(\{f_i\}\): this follows from the linearity of the polynomials of \(\{f_i\}-\{g_i\}\) and the fact that every \(a\in A\) is a common solution (by definition of \(A\)). Thus, it suffices to show the result for the set \(\{g_i\}\).

Next, consider the polynomial \(g\in\F_2[x_1,\dots,x_n]\) defined as
\[
g(x):=\prod_i(1+g_i(x)).
\]
It is straightforward to verify that \(g\) has degree at most \(\Delta\), and that \(g(x)=1\) if and only if \(x\) is a common solution to \(\{g_i\}\). Now, suppose for contradiction that \(A+A\) contains no nontrivial common solution to \(\{g_i\}\): that is, for every distinct \(x,y\in A\) it holds that \(g(x+y)=0\). Then, consider the \(2^n\times2^n\) matrix \(M\) with entries \(M_{x,y}=g(x+y)\) for every \(x,y\in\F_2^n\). Define \(k:=|A|\), and let \(M[A,A]\) denote the \(k\times k\) submatrix of \(M\) obtained by taking the rows and columns of \(M\) indexed by \(A\). Since \(0\) is a common solution to \(\{g_i\}\), we get that \(M[A,A]=I_k\) and thus
\[
\mathsf{rank}(M)\geq\mathsf{rank}(M[A,A])=\mathsf{rank}(I_k)=k>2\binom{n}{\lfloor \Delta/2\rfloor},
\]
which directly contradicts \cref{lem:clp}.
\end{proof}

\subsubsection{Proofs of its corollaries}\label{subsubsec:proof-low-wt-cw:corollaries}

Now that we have proven our general low-weight Chevalley-Warning theorem (\cref{thm:low-wt-cw:technical}), we are ready to prove its corollaries. These proofs are relatively straightforward, but somewhat tedious. We start with the proof of \cref{cor:low-wt-cw:main-technical-cor}, which says that we can always find a nontrivial common solution of Hamming weight \(w\leq 8\Delta + 8D/\log(n/D)+8\).

\begin{proof}[Proof of \cref{cor:low-wt-cw:main-technical-cor}]
First, note that the result is true if \(8\Delta+8D/\log(n/D)+8\geq D+\Delta+1\), as the existence of a nontrivial common solution with Hamming weight \(\leq D+\Delta+1\) is immediate from Chevalley-Warning (see the discussion following \cref{thm:absolutely-standard-Chevalley-Warning}). Thus we henceforth assume, without loss of generality, that
\begin{align}\label{eq:excellent-inequality}
8\Delta+8D/\log(n/D)+8<\Delta+D+1\leq n,
\end{align}
where the last inequality comes from the given hypothesis \(D+\Delta<n\). This string of inequalities will come in handy later.

Now, by \cref{thm:low-wt-cw:technical}, we know that for any \(W\) satisfying both
\begin{align}
    \binom{n}{\leq\lfloor W/2\rfloor}&>2^{(D+\Delta+1)\cdot2},\label{eq:first-cw}\\
    \binom{n}{\leq\lfloor W/2\rfloor}&>\binom{n}{\leq\lfloor\Delta/2\rfloor}^2,\label{eq:second-cw}
    \end{align}
    it holds that there is a nontrivial common solution of Hamming weight \(\leq W\). We seek to find the smallest \(W\) satisfying both \cref{eq:first-cw,eq:second-cw}.

    We start with \cref{eq:second-cw}. First, by relatively straightforward binomial inequalities, observe
    \[
    \binom{n}{\leq\lfloor\Delta/2\rfloor}^2\leq\binom{2n}{\leq 2\lfloor\Delta/2\rfloor}\leq\binom{n}{\leq4\lfloor\Delta/2\rfloor},
    \]
    where the last inequality uses the fact that \(8\Delta<n\) (as implied by \cref{eq:excellent-inequality}). Thus, any \(W\) satisfying
    \begin{align*}
        \binom{n}{\leq\lfloor W/2\rfloor}>\binom{n}{\leq 4\lfloor\Delta/2\rfloor}
    \end{align*}
    also satisfies \cref{eq:second-cw}. And the above inequality is satisfied by any \(W\) satisfying
    \[
    \lfloor W/2\rfloor>4\lfloor\Delta/2\rfloor,
    \]
    where we have again used the assumption that \(8\Delta<n\) to ensure \(4\lfloor\Delta/2\rfloor<n\). Thus, any \(W\) satisfying
    \[
    W>4\Delta+1.
    \]
    also satisfies \cref{eq:second-cw}.

   We now consider \cref{eq:first-cw}. By \cref{eq:excellent-inequality}, we may assume \(D>7\Delta+7\), which means that \(\Delta<D/7-1\) and thus \((D+\Delta+1)\cdot2<16D/7\). As a result, any \(W\) satisfying
   \begin{align}\label{eq:tough-inequality}
   \binom{n}{\leq\lfloor W/2\rfloor}>2^{16D/7}
   \end{align}
   also satisfies \cref{eq:first-cw}. We consider two cases.
   
   \emph{Case (i): \(D\leq \log n\):} By \cref{eq:excellent-inequality}, we may assume \(n>8\). It is now easy to verify that for any \(W\geq 8\),
   \[
   \binom{n}{\leq\lfloor W/2 \rfloor}\geq\binom{n}{\leq 4} > n^{16/7}\geq 2^{16D/7},
   \]
   thereby satisfying \cref{eq:tough-inequality}.
   
   \emph{Case (ii): \(D>\log(n)\):} Let \(1\leq k\leq n\) be a parameter we will pick later, and notice that for any \(W\geq 2k\) we have
   \[
   \binom{n}{\leq\lfloor W/2\rfloor}\geq\binom{n}{\leq\lfloor k\rfloor}\geq\binom{n}{\lfloor k\rfloor}\geq\left(\frac{n}{\lfloor k\rfloor}\right)^{\lfloor k\rfloor}.
   \]
   Thus if \(1\leq k\leq n\) is a parameter satisfying \(\lfloor k\rfloor\log(n/\lfloor k\rfloor)>16D/7\)
   then any \(W\geq 2k\) satisfies \cref{eq:tough-inequality}. Now, define \(k:=4D/\log(n/D)\). By the case condition and \cref{eq:excellent-inequality}, we have \(4\leq k<n\), and thus
   \[
   3D/\log(n/D)\leq\lfloor k\rfloor\leq 4D/\log(n/D).
   \]
   By \cref{eq:excellent-inequality}, we may assume \(\log(n/D)\geq 8\), and thus
   \[
   \lfloor k\rfloor \log(n/\lfloor k\rfloor)\geq\frac{3D}{\log(n/D)}\log\left(\frac{n\log(n/D)}{4D}\right)\geq\frac{3D}{\log(n/D)}\log\left(\frac{2n}{D}\right)\geq3D > 16D/7.
   \]
   Thus, every \(W\geq2k=8D/\log(n/D)\) satisfies \cref{eq:tough-inequality}.
   
   To conclude, we get that every
   \[
   W\geq\max\{4\Delta+2,8,8D/\log(n/D)\}
   \]
   satisfies both \cref{eq:first-cw,eq:second-cw}, which completes the proof.
\end{proof}

We now turn towards proving \cref{cor:projection-cw}, which says that, not only is it possible to find a low weight common solution, but it is even possible to force this solution to be supported on a target set of coordinates \(S\). The proof follows quite directly by combining \cref{cor:low-wt-cw:main-technical-cor} with the idea of function restrictions.

\begin{proof}[Proof of \cref{cor:projection-cw}]
For any \(y\in\F_2^{|S|}\), let \(y^+\in\F_2^n\) denote the unique string where \(y^+_S=y\) and \(y^+_{\overline{S}}=0^{n-|S|}\). Now, for each \(f_i\) let \(g_i\in\F_2[x_1,\dots,x_{|S|}]\) denote the polynomial \(g_i(x):=f_i(x^+)\). Notice that each \(\deg(g_i)\leq\deg(f_i)\). Now, for each \(i\) create a polynomial \(h_i\in\F_2[x_1,\dots,x_{|S|}]\) as follows. If \(\deg(f_i)>1\) but \(\deg(g_i)=1\): assume without loss of generality that \(g_i(x)\neq x_1\), and set \(h_i(x):=1+(1+x_1)(1+g_i(x))\); otherwise, just define \(h_i(x):=g_i(x)\).

Consider the set of polynomials \(\{h_i\}\). Notice that the nonlinear degree of \(\{h_i\}\) is at most \(\Delta\), and the linear degree of \(\{h_i\}\) is at most \(D\). Thus, by \cref{cor:low-wt-cw:main-technical-cor}, there is a nontrivial common solution \(y\) to \(\{h_i\}\) with weight \(w\leq8\Delta+8D/\log(|S|/D)+8\). It is then straightforward to verify that \(y\) must be a common solution to \(\{g_i\}\), and that \(y^+\) must be a common solution to \(\{f_i\}\). Furthermore, \(y^+\) is clearly supported on \(S\) and has the same Hamming weight as \(y\).
\end{proof}

\subsection{A local version of Cohen-Tal}\label{subsec:cohen-tal-local-sec}

Equipped with our low-weight Chevalley-Warning theorem, we are now ready to prove our local version of Cohen-Tal (\cref{thm:local-cohen-tal:restated}). Recall that this result says every degree \(\leq r\) polynomial admits a large local subspace on which it is constant, and that the lower bounds that we get on its size are tight. For convenience, we split this theorem into two lemmas: one which claims the lower bounds (\cref{lem:lower-bound-cohen-tal-us}), and one which claims their tightness (\cref{lem:upper-bound-cohen-tal-us}). We prove these lemmas in \cref{subsubsec:the-lower-bounds,subsubsec:tightness-of-the-lower-bounds}, respectively. \cref{thm:local-cohen-tal:restated} will then follow immediately.

\subsubsection{The lower bounds}\label{subsubsec:the-lower-bounds}

We start with a proof of the lower bounds in \cref{thm:local-cohen-tal:restated}, which is much more challenging to prove than the tightness result.

\begin{lemma}[Lower bound of \cref{thm:local-cohen-tal:restated}]\label{lem:lower-bound-cohen-tal-us}
There exists a universal constant \(c>0\) such that for every \(n,r,d\in\N\) such that \(2\leq r\leq c\log(n)\) and \(d\leq n^{\frac{1}{r-1}-2^{-10r}}/\log n\), the following holds. For any degree \(\leq r\) polynomial \(f\in\F_2[x_1,\dots,x_n]\), there exists a \(d\)-local affine subspace \(X\subseteq\F_2^n\) of dimension
\[
k\geq cr (d n\log n)^{1/r}
\]
on which \(f\) is constant.
\end{lemma}

Our proof of \cref{lem:lower-bound-cohen-tal-us} will combine our low-weight Chevalley-Warning theorem (\cref{cor:projection-cw}) with a (known) ingredient from Cohen and Tal \cite{CT15} that involves \emph{directional derivatives}. Given a polynomial $f\in\F_2[x_1,\dots,x_n]$ and a set of vectors \(S\subseteq\F_2^n\), we let \(f_S\) denote the \emph{derivative of \(f\) in the directions of \(S\)}, where
\begin{equation*}
    f_S(x):=\sum_{T\subseteq S} f\left(x+\sum_{v\in T}v\right).
\end{equation*}
One useful observation is that \(\deg(f_S)\leq\max\{0,\deg(f)-|S|\}\). In addition to this, the key ingredient about directional derivatives that we import from Cohen and Tal is the following.

\begin{lemma}[\hspace{1sp}\cite{CT15}]\label{lem:derivatives-yahoo}
For any degree \(\leq r\) polynomial \(f\in\F_2[x_1,\dots,x_n]\) and \(B\subseteq\F_2^n\),
\[
f\left(x + \sum_{v\in S}v \right)=0\text{ for all }S\subseteq B\iff f_S(x)=0\text{ for all }S\subseteq B\text{ of size }|S|\leq r.
\]
\end{lemma}

This shows that \(f\) is \(0\) on the entire affine space \(x+\mathsf{span}(B)\) if and only if \(0\) is a common solution to all \((\leq r)\)-wise directional derivatives of \(f\) across B. Now, given \cref{lem:derivatives-yahoo} and our low-weight Chevalley-Warning theorem (\cref{cor:projection-cw}),  we are ready to prove our lower bound for extracting from \(d\)-local affine sources (\cref{lem:lower-bound-cohen-tal-us}).

\begin{proof}[Proof of \cref{lem:lower-bound-cohen-tal-us}]
Assume without loss of generality that \(f(0)=0\), for otherwise we can work with the polynomial \(1+f\). We will iteratively build a basis \(B\subseteq\F_2^n\) of a \(d\)-local subspace \(X\subseteq\F_2^n\) on which \(f\) is constantly \(0\).

Towards this end, we start by initializing several sets:
\begin{itemize}
    \item \(B\gets\emptyset\).
    \item \(\mathsf{UNIQUE}\gets\emptyset\).
    \item \(\mathsf{SATURATED}\gets\emptyset\).
    \item \(P\gets\{f\}\).
\end{itemize}
Now, \textbf{while} there exists a common nontrivial solution \(b\in\F_2^n\) to the set \(P\subseteq\F_2[x_1,\dots,x_n]\) such that \(b\) is supported on \([n]\setminus(\mathsf{UNIQUE}\cup\mathsf{SATURATED})\), \textbf{do} the following:
\begin{itemize}
    \item Let \(b^\ast\) be the lowest (Hamming) weight vector of this type (breaking ties arbitrarily).
    \item Set \(B\gets B\cup\{b^\ast\}\).
    \item Define \(\alpha\in[n]\) as the smallest index such that \(b^\ast_\alpha=1\).
    \item Set \(\mathsf{UNIQUE}\gets\mathsf{UNIQUE}\cup\{\alpha\}\).
    \item Set \(\mathsf{SATURATED}\gets\{i\in[n] : \text{ there are \(d\) vectors }v\in B\text{ with }v_i=1\}\).
    \item For each \(S\subseteq B\) of size \(|S|\leq r\), set \(P\gets P\cup\{f_S\}\).
\end{itemize}
This completes the construction of \(B\).

We now proceed with the analysis. First, we argue that at the end of the above construction, \(B\) will hold a basis for a \(d\)-local subspace \(X:=\mathsf{span}(B)\). To see why, first define \(T\) to be the number of times the above loop executes, and for each \(i\in[T]\) let \(b^i\in\F_2^n\) denote the vector added to \(B\) in iteration \(i\). Then, let \(\alpha(i)\in[n]\) denote the smallest index such that \(b^i_{\alpha(i)}=1\), and observe that \(b^j_{\alpha(i)}=0\) for all \(j>i\), because of bullets 3-4 above (combined with the loop condition). Thus \(B=\{b^1,\dots,b^T\}\) is a collection of linearly independent vectors: in other words, a basis. Furthermore, bullet 5 above (combined with the loop condition) ensures that \(X:=\mathsf{span}(B)\) is \(d\)-local.

Thus we have argued that \(X\) is a \(d\)-local subspace of dimension \(T\). Now, by combining bullet 6 above with the loop condition and \cref{lem:derivatives-yahoo}, we immediately get that \(f(x)=0\) for all \(x\in X\). Thus, all that remains is to get a lower bound on \(T\), the number of times that the loop executes.

We call the first iteration of the loop \emph{iteration \(1\)}, and we call the pseudocode that precedes the while loop \emph{iteration \(0\)}. Now, for all \(t=0,\dots,T\), we define the following variables, for convenience:

\begin{itemize}
    \item Let \(D_t\) denote the degree of the system \(P\) upon completing iteration \(t\).
    \item Let \(\Delta_t\) denote the nonlinear degree of the system \(P\) upon completing iteration \(t\).
    \item Let \(u_t\) denote the size of \(|\mathsf{UNIQUE}|\) upon completing iteration \(t\).
    \item Let \(s_t\) denote the size of \(|\mathsf{SATURATED}|\) upon completing iteration \(t\).
    \item Let \(w_t\) denote the Hamming weight of the vector \(b^\ast\) selected in iteration \(t\).
\end{itemize}
Note that \(w_0\) is undefined above, and for convenience we define \(w_0:=0\). Furthermore, for any \(t>T\), it will be convenient to define \(D_t := D_T, \Delta_t := \Delta_T\), and so on. Now, before we turns towards getting a lower bound on \(T\), we \emph{upper bound} the above quantities.

\begin{itemize}
\item \(D_t\leq\sum_{i=0}^{r-1}\binom{t}{i}(r-i)\leq r\binom{t}{\leq r-1}\). This is because, at the end of iteration \(t\) we have: \(|B|=t\); and \(P\) holds the set of polynomials \(f_S\) for all \(S\subseteq B\) of size \(|S|\leq r\); and \(\deg(f_S)\leq\max\{0,\deg(f)-|S|\}\).
\item \(\Delta_t\leq\sum_{i=0}^{r-2}\binom{t}{i}(r-i)\leq r\binom{t}{\leq r-2}\). This holds for the same reasons as above, except that we are now ignoring the polynomials \(f_S\) for all \(S\subseteq B\) of size \(|S|\geq r-1\) (since these have degree at most \(1\)).
\item \(u_t=t\), since we grow \(\mathsf{UNIQUE}\) by a (unique) coordinate in every iteration.
\item \(s_t\leq\frac{1}{d}\sum_{i=1}^tw_i\), since otherwise the sum of the Hamming weights of the vectors \(b^\ast\) selected in iterations \(1,\dots,t\) is at least \(ds_t>\sum_{i=1}^tw_i\), a contradiction.
\item \(w_t\leq 8\Delta_{t-1}+8D_{t-1}/\log(\frac{n-{u_{t-1}-s_{t-1}}}{D_{t-1}})+8\), by \cref{cor:projection-cw}.
\end{itemize}
We now turn towards getting a lower bound on \(T\). Here, the key observation is that it \emph{must} hold that
\[
D_T \geq n - (u_T + s_T),
\]
because otherwise the degree of the set of polynomials is strictly less than the size of the set on which it must be supported, so \cref{cor:projection-cw} trivially implies that there is \emph{some} nontrivial common solution supported on \(S\) (thereby forcing the next execution of the while condition to pass). A straightforward consequence of this is \(T\geq1\) since \(D_0 = r < n=n-(u_0+s_0)\).

Instead of getting a lower bound on \(T\), it will facilitate the analysis to instead lower bound some \(\tau\leq T\). In particular, let \(\tau\in\{0,1,\dots,T-1\}\) be the smallest integer such that
\begin{align}\label{eq:geqnover2}
D_{\tau+1}+u_{\tau+1}+s_{\tau+1}\geq n/2.
\end{align}
Such a \(\tau\) must exist, since we saw above that \(D_T+u_T+s_T\geq n\). We now seek to lower bound \(\tau\), which will in turn lower bound \(T\).

We start with a basic (but useful) lower bound on \(\tau\). In particular, we claim that \(\tau\geq2r\). To see why, suppose for contradiction \(\tau<2r\). Then by definition of \(\tau\) it holds that \(D_{2r}+u_{2r}+s_{2r}\geq n/2\). But by our upper bounds on these quantities (and the fact that \(r\geq 1\)) we have
\begin{align*}
D_{2r}+u_{2r}+s_{2r}&\leq D_{2r} + u_{2r} + 8\sum_{i=0}^{2r-1}(\Delta_i + D_i + 1)\\
&\leq D_{2r}+u_{2r} + 16r\cdot(\Delta_{2r} + D_{2r}+1)\\
&\leq r\cdot\binom{2r}{\leq r-1} + 2r + 16r\cdot(r\cdot\binom{2r}{\leq r-2} + r\cdot\binom{2r}{\leq r-1}+r)\\
&\leq r\cdot 2^{2r} + 2r + 16r\cdot(r\cdot 2^{2r} + r\cdot 2^{2r}+r)\\
&\leq 2^{12r}.
\end{align*}
Thus we get that \(n/2\leq D_{2r}+u_{2r} + s_{2r}\leq 2^{12r}\), which implies that \(r\geq \log(n)/13\). But the theorem hypothesis claims that \(r\leq c\log(n)\) for some universal constant \(c\); thus, we can just pick \(c\) later to ensure that \(c<\frac{1}{13}\), which would yield a contradiction and conclude the proof of the claim that \(\tau\geq2r\). And since \(r\geq1\) by the theorem hypothesis, we also know that \(\tau\geq2\).

Given the above, we now proceed to get a more general lower bound on \(\tau\). We will do so by sandwiching the quantity \(D_{\tau+1}+u_{\tau+1}+s_{\tau+1}\) between two inequalities. Towards this end, the key observation is that for any \(i\in[\tau]\), it holds that \(u_i+s_i\leq u_i+s_i+D_i < n/2\).\footnote{This useful inequality is the reason we work with \(\tau\) instead of \(T\), as this is not necessarily true for any \(i\in[T]\).} Combining this with \cref{eq:geqnover2} and our earlier upper bounds on \(s_t,w_t\), we have
\begin{align*}
n/2\leq D_{\tau+1}+u_{\tau+1}+s_{\tau+1}&\leq D_{\tau+1}+u_{\tau+1}+\frac{1}{d}\sum_{i=1}^{\tau+1}w_i\\
&\leq D_{\tau+1}+u_{\tau+1}+\frac{8}{d}\sum_{i=0}^{\tau}\left(\Delta_i+\frac{D_i}{\log\left(\frac{n-u_i-s_i}{D_i}\right)}+1\right)\\
&\leq D_{\tau+1}+u_{\tau+1} + \frac{8}{d}\sum_{i=0}^\tau\left(\Delta_\tau + \frac{D_\tau}{\log\left(\frac{n-u_\tau-s_\tau}{D_\tau}\right)}+1\right)\\
&\leq D_{\tau+1}+(\tau+1) + \frac{8}{d}(\tau+1)\left(\Delta_\tau + \frac{D_\tau}{\log\left(\frac{n-u_\tau-s_\tau}{D_\tau}\right)}+1\right)\\
&\leq D_{\tau+1}+(\tau+1) + \frac{8}{d}(\tau+1)\left(\Delta_\tau + \frac{D_\tau}{\log\left(\frac{n}{2D_\tau}\right)}+1\right)\\
&\leq D_{\tau+1} + 2\tau + \frac{16\tau}{d}\left(\Delta_\tau+\frac{D_\tau}{\log\left(\frac{n}{2D_\tau}\right)}+1\right)\\
&\leq D_{\tau+1} + 18\tau + \frac{16\tau}{d}\left(\Delta_\tau + \frac{D_\tau}{\log\left(\frac{n}{2D_\tau}\right)}\right).
\end{align*}
Thus we know that if we define
\begin{align*}
K_1 &:= D_{\tau+1}+18\tau + \frac{16\tau}{d}\Delta_\tau,\\
K_2 &:= \frac{16\tau D_\tau}{d\log(\frac{n}{2D_\tau})},
\end{align*}
it must hold that \(K_1 + K_2\geq n/2\), or rather we know that either \(K_1\geq n/4\) or \(K_2\geq n/4\) must hold. We analyze these cases separately, and get a lower bound on \(\tau\) in each case. But before we do so, it will be useful to recall that for any integers \(b>a\geq0\) it holds that \(\binom{b}{a}\leq\binom{b}{a+1}\) if \(a+1\leq b/2\). Since we saw earlier that \(\tau\geq2r\), we have \(\binom{2\tau}{\leq r-1}\leq r\cdot\binom{2\tau}{r-1}\) and \(\binom{\tau}{\leq r-2}\leq r\cdot\binom{\tau}{r-2}\) and \(\binom{\tau}{\leq r-1}\leq r\cdot \binom{\tau}{r-1}\). We now proceed with the case analysis.

\emph{Case (1): \(K_1\geq n/4\)}: By applying the upper bounds on \(D_t,\Delta_t\) obtained earlier in the proof, the standard binomial inequality \(\binom{n}{k}\leq(en/k)^k\), and the fact that \(r\geq2\), we have:
\begin{align*}
    n/4 \leq K_1 = D_{\tau+1}+18\tau + \frac{16\tau}{d}\Delta_\tau &\leq r\cdot\binom{\tau+1}{\leq r-1}+18\tau + \frac{16\tau}{d}r\cdot\binom{\tau}{\leq r-2}\\
    &\leq r\cdot\binom{2\tau}{\leq r-1}+18\tau+\frac{16\tau}{d}r\cdot\binom{\tau}{\leq r-2}\\
    &\leq r^2\cdot\binom{2\tau}{r-1}+18\tau+\frac{16\tau r^2}{d}\cdot\binom{\tau}{r-2}\\
    &\leq r^2\cdot(4e)^{r-1}\cdot\left(\frac{\tau}{r}\right)^{r-1} + 18r\left(\frac{\tau}{r}\right)^{r-1}+\frac{16}{d}r^3\cdot (3e)^r\cdot\left(\frac{\tau}{r}\right)^{r-1}\\
    &\leq 3\cdot18\cdot(4e)^r\cdot r^3\cdot\left(\frac{\tau}{r}\right)^{r-1}\\
    &\leq 2^6\cdot2^{7r}\cdot\left(\frac{\tau}{r}\right)^{r-1},
\end{align*}
which implies that
\[
n\leq 2^8\cdot2^{7r}\cdot\left(\frac{\tau}{r}\right)^{r-1}\leq 2^{15r}\cdot\left(\frac{\tau}{r}\right)^{r-1},
\]
which gives
\[
\tau\geq r\cdot\left(n\cdot2^{-15r}\right)^{\frac{1}{r-1}}\geq2^{-30}\cdot rn^{\frac{1}{r-1}}.
\]

\emph{Case (2): \(K_2\geq n/4\)}: In this case we have
\[
n/4\leq K_2 = \frac{16\tau D_\tau}{d\log\left(\frac{n}{2D_\tau}\right)},
\]
which of course implies
\[
64\tau D_\tau\geq dn\log\left(\frac{n}{2D_\tau}\right).
\]
Now, notice that our earlier upper bounds on \(D_\tau\) yield
\[
D_\tau\leq r\cdot\binom{\tau}{\leq r-1}\leq r^2\cdot\binom{\tau}{r-1}\leq r^2\cdot(2e)^{r-1}\cdot\left(\frac{\tau}{r}\right)^{r-1}\leq 2^{5r}\cdot\left(\frac{\tau}{r}\right)^{r-1},
\]
and combining this with the previous inequality yields
\[
64\tau\cdot 2^{5r}\cdot\left(\frac{\tau}{r}\right)^{r-1}\geq 64\tau D_\tau\geq dn\log\left(\frac{n}{2D_\tau}\right)\geq dn\log\left(\frac{n}{2^{5r+1}\cdot\left(\frac{\tau}{r}\right)^{r-1}}\right).
\]
Then, applying straightforward bounds on both sides of the above chain of inequalities yields
\begin{align}\label{eq:almost-done-with-hard-direction}
2^{12r}\cdot\left(\frac{\tau}{r}\right)^r\geq dn\log\left(\frac{n}{2^{6r}\cdot\left(\frac{\tau}{r}\right)^{r-1}}\right).
\end{align}
We now seek to get a lower bound on all \(\tau\) that satisfy the above inequality. Towards this end, notice that for any \(\alpha>0\) such that
\begin{align}\label{eq:actually-almost-done-with-hard-direction}
2^{12r}\cdot\left(\frac{\alpha}{r}\right)^r< dn\log\left(\frac{n}{2^{6r}\cdot\left(\frac{\alpha}{r}\right)^{r-1}}\right),
\end{align}
it holds that all \(\tau\) that satisfy \cref{eq:almost-done-with-hard-direction} must also satisfy \(\tau\geq\alpha\): this is because as \(\alpha\) decreases, the left hand side of the above inequality decreases, while its right hand side increases.

It is now straightforward to verify that
\[
\alpha=2^{-30}\cdot r\cdot \left(dn\log n\right)^{1/r}
\]
satisfies \cref{eq:actually-almost-done-with-hard-direction}, as long as \(d\log n\leq n^{\frac{1}{r-1} - 2^{-10r}}\). Thus in this case we have
\[
\tau\geq\alpha = 2^{-30}\cdot r\cdot (dn\log n)^{1/r},
\]
provided \(d\log n\leq n^{\frac{1}{r-1}-2^{-10r}}\).

To conclude, since one of the cases must hold, we get that
\[
\tau\geq 2^{-30}\cdot\min\{rn^{\frac{1}{r-1}},r(dn\log n)^{1/r}\}
\]
if \(d\log n\leq n^{\frac{1}{r-1}-2^{-10r}}\). But given this condition on \(d\log n\), it always holds that \(r(dn\log n)^{1/r}\leq rn^{\frac{1}{r-1}}\). Thus, as long as \(d\log n\leq n^{\frac{1}{r-1} - 2^{-10r}}\), we get that
\[
T\geq \tau\geq 2^{-30}r(dn\log n)^{1/r},
\]
as desired.
\end{proof}

\subsubsection{On the tightness of our lower bounds}\label{subsubsec:tightness-of-the-lower-bounds}

At last, all that remains is to show our lower bounds (in \cref{lem:lower-bound-cohen-tal-us}) are tight, in the following sense.

\begin{lemma}[Upper bound of \cref{thm:local-cohen-tal:restated}]\label{lem:upper-bound-cohen-tal-us}
There exist universal constants \(C,c>0\) such that a random degree \(\leq r\) polynomial \(f\sim\F_2[x_1,\dots,x_n]\) is an extractor for \(d\)-local affine sources with min-entropy
\[
k\geq Cr\cdot(dn\log n)^{1/r}
\]
and error \(\eps=2^{-c k/r}\), except with probability at most \(2^{-c{k\choose\leq r}}\) over the selection of \(f\).
\end{lemma}

Previously, we showed (in \cref{sec:new-upper-bounds}) that low-degree polynomials extract from \(d\)-local sources (\cref{thm:localextintro:restated}), and our main ingredients were: (i) a reduction from \(d\)-local sources to \(d\)-local NOBF sources; and (ii) a result showing that low-degree polynomials extract from \(d\)-local NOBF sources (which relied heavily on a certain correlation bound we proved in \cref{subsubsec:correlation-bounds-sec}).

To prove \cref{lem:upper-bound-cohen-tal-us}, we wish to show that low-degree polynomials extract from \(d\)-local \emph{affine} sources. These are much more structured than general \(d\)-local sources, and we are able to optimize the above framework to skip step (i) and rely on a stronger known correlation bound in step (ii). In particular, our key ingredient will be the following, which shows that a random low-degree polynomial has small bias (i.e., small correlation with the constant \(1\) function).

\begin{lemma}[\hspace{1sp}\cite{low-bias-polys}]\label{lem:low-bias-poly}
For any fixed \(\eps>0\) there exist \(0<c_1,c_2<1\) such that the following holds. Let \(f\sim\F_2[x_1,\dots,x_n]\) be a random polynomial of degree \(\leq r\) where \(r\leq (1-\eps)n\). Then
\[
\Pr[\mathsf{bias}(f)>2^{-c_1n/r}]\leq 2^{-c_2{n\choose\leq r}}.
\]
\end{lemma}

Given this result, we are now able to prove \cref{lem:upper-bound-cohen-tal-us} without too much trouble.

\begin{proof}[Proof of \cref{lem:upper-bound-cohen-tal-us}]

Let \(\X\sim\F_2^n\) be a \(d\)-local affine source of dimension \(k\). It is straightforward to show, via Gaussian elimination, that the following holds: there exists a subset \(S\subseteq[n]\) of size \(k\) such that \(\X_S\) is uniform over \(\F_2^k\), and for every other \(j\in[n]\) there exists an affine function \(\ell_j:\F_2^k\to\F_2\) such that \(\X_j=\ell_j(\X_S)\). Without loss of generality, assume that \(S=[k]\). Thus if we consider the function \(\ell:\F_2^k\to\F_2^n\) defined as
\[
\ell(y_1,\dots,y_k):=(y_1,\dots,y_k,\ell_{k+1}(y_1,\dots,y_k),\dots,\ell_n(y_1,\dots,y_k))
\]
we of course have \(\X=\ell(\U_k)\), where \(\U_k\sim\F_2^k\) denotes the uniform random variable.

Now, let \(f\sim\F_2[x_1,\dots,x_n]\) be a random polynomial of degree \(\leq r\). Notice that
\[
|f(\X)-\U_1|=|f(\ell(\U_k))-\U_1|=\mathsf{bias}(f(\ell))/2
\]
So we want to upper bound the probability that \(\mathsf{bias}(f(\ell))\) is large, over our random selection of \(f\). Well, consider fixing whether each monomial \(x^S, S\not\subseteq[k]\) exists. Notice that under any such fixing, \(f(\ell)\) actually turns into a polynomial of the form \(f^\pr + g\), where \(f^\pr\sim\F_2[x_1,\dots,x_n]\) is a uniformly random degree \(\leq r\) polynomial, and \(g\) is a sum of monomials, each of size at most \(r\), where each term in each monomial is linear (i.e., degree \(1\)). In other words, \(g\) is a fixed degree \(\leq r\) polynomial, and thus \(f^\pr+g\) is a uniformly random degree \(\leq r\) polynomial over \(k\) variables.

Thus, by \cref{lem:low-bias-poly}, we get that under \emph{every} such fixing, \(\bias(f(\ell))\leq 2^{-c_1k/r}\) except with probability \(\leq 2^{-c_2{k\choose\leq r}}\). Thus for a given \(d\)-local affine source \(\X\sim\F_2^n\) of min-entropy \(k\), it holds that a random degree \(\leq r\) polynomial extracts from \(\X\) with error \(\leq 2^{-c_1k/r}\), except with probability \(\leq 2^{-c_2{k\choose\leq r}}\).

Now, it is not hard to show that there are at most \({k\choose\leq d}^n\cdot2^n\) \(d\)-local affine sources of dimension \(k\). Thus, by a union bound, we know that a random degree \(\leq r\) polynomial extracts from \emph{all} \(d\)-local affine sources \(\X\sim\F_2^n\) of min-entropy \(k\), except with probability at most
\[
2^{-c_2{k\choose \leq r}}\cdot{k\choose\leq d}^n\cdot 2^n.
\]
Thus as long as \({k\choose\leq d}^n\cdot 2^n\leq 2^{c_2{k\choose\leq r}/2}\), it holds that a random degree \(\leq r\) polynomial extracts from \(d\)-local affine sources with min-entropy \(k\) with error \(\leq 2^{-c_1 k/r}\), with probability \(\geq 1-2^{-c_2{k\choose\leq r}/2}\). It is now easy to verify that the inequality \({k\choose\leq d}^n\cdot2^n\leq 2^{c_2{k\choose\leq r}/2}\) holds for the stated lower bound on \(k\).
\end{proof}

\dobib

\section{A barrier}

In this final section, we prove a barrier result, which shows that previous techniques cannot be used to extract from local sources with min-entropy below \(\sqrt{n}\).

In more detail, previous techniques \cite{DW12,Vio14} show that local sources are close to a convex combination of affine sources (with almost the same min-entropy), which means that a black-box affine extractor can also extract from local sources. However, the arguments in \cite{DW12,Vio14} only work if the local sources start out with min-entropy \(\gg\sqrt{n}\). On the other hand, it is not completely clear if this is an artifact of the exact arguments made in those works, or if local sources with min-entropy below \(\sqrt{n}\) simply cannot be shown to be close to a convex combination of affine sources, using \emph{any} argument.

In our barrier result, stated below, we show that the latter is the case.

\begin{theorem}[\cref{claim:barrierintro}, formal version]\label{thm:barrier:technical-section}
There exists a \(2\)-local source \(\X\sim\zo^n\) with min-entropy \(k\geq\sqrt{n}\) that is \((1/2)\)-far from a convex combination of affine sources with min-entropy \(5\).
\end{theorem}

This barrier result confirms that it is impossible to use affine extractors in a standard black-box manner (i.e., using convex combinations) to extract from local sources with min-entropy below \(\sqrt{n}\), for two reasons:

\begin{enumerate}
    \item The statistical distance to a convex combination of affine sources is too large: since the statistical distance in \cref{thm:barrier:technical-section} is \(\geq1/2\), a black-box application of an affine extractor result in error \(\geq1/2\), which is no better than the error of the trivial extractor (which just outputs a constant).
    \item The entropy of the affine sources in the convex combination is too low: the entropy of the affine sources in \cref{thm:barrier:technical-section} is \(5\), which is much less than the \(\Omega(\log n)\) entropy required for an affine extractor to even exist \cite{AggarwalBGS21}.
\end{enumerate}

In order to prove \cref{thm:barrier:technical-section}, we use a \(2\)-local source \(\X\sim\zo^n\) called the \emph{clique source}. We define the clique source over \(n=k+\binom{k}{2}\) bits to be the random variable \(\X\sim\zo^n\) generated by sampling \(k\) independent and uniform bits \(\Y_1,\dots,\Y_k\) and outputting
\[
\X=(\Y_1,\dots,\Y_k, (\Y_i\cdot\Y_j)_{1\leq i < j \leq k}).
\]
Then, we prove the following, from which \cref{thm:barrier:technical-section} is immediate.

\begin{theorem}\label{thm:far-convex-combo}
	Let \(\X\sim\zo^n\) be the clique source over \(n={k+1\choose2}\) bits. Then for any \(t\), it holds that \(\X\) is \((1-\eps)\)-far from a convex combination of affine sources with min-entropy \(t\), where
	\[
	\eps=2^{-(t-3)/2}.
	\]
\end{theorem}

The main lemma we use to prove \cref{thm:far-convex-combo} roughly shows that the collection of all cliques in the space of \(n\)-vertex graphs is a \emph{subspace-evasive set} (a well-studied object in pseudorandomness: see, e.g., \cite{DL12}). More formally, fix any \(n\in\N\) and let \(m:=n+{n\choose 2}\). We identify \(\F_2^m\) with the set of all \(n\)-vertex (undirected) graphs as follows. First, fix an arbitrary identification of \([m]\) with \([n] \cup {[n]\choose2}\), and let \(x\in\F_2^m\) represent the graph \(G_x=(V_x,E_x)\) where \(i \in V_x\) if and only if \(x_i = 1\) and \(e\in E_x\) if and only if \(x_e=1\). Then, we say that \(x\in\F_2^m\) is a clique if (and only if) \(E_x={V_x\choose2}\), and prove the following.

\begin{lemma}\label{lem:main-subspace-evasive}
Let \(\mathcal{Q}\subseteq\F_2^m\) denote the collection of all cliques. Then for any affine subspace \(S\subseteq\F_2^m\) of dimension \(k\),
	\[
	\frac{|\mathcal{Q}\cap S|}{|S|}\leq2^{-(k-3)/2}.
	\]
	That is, an exponentially small fraction of graphs in \(S\) are cliques.
\end{lemma}

Before we prove \cref{lem:main-subspace-evasive} (in the following subsection), we show how it implies \cref{thm:far-convex-combo}.

\begin{proof}[Proof of \cref{thm:far-convex-combo}]
Let \(\Y\sim\zo^n\) be a convex combination of affine sources \(\{\Y^{(i)}\}_i\), each with min-entropy \(t\). Let \(S=\supp(\X)\) and \(\overline{S}:=\zo^n-S\). We have
\begin{align*}
|\X-\Y|&\geq\Pr[\Y\in\overline{S}]-\Pr[\X\in\overline{S}]=\Pr[\Y\in\overline{S}]\\
&\geq\min_i\Pr[\Y^{(i)}\in\overline{S}]=1-\max_i\Pr[\Y^{(i)}\in S]\\
&=1-\max_i\frac{|\supp(\X)\cap\supp(\Y^{(i)})|}{|\supp(\Y^{(i)})|}.
\end{align*}
It is easy to verify that each element \(x\in\supp(\X)\) is a clique, and thus by \cref{lem:main-subspace-evasive} the above equality is at least \(1-2^{-(t-3)/2}\), as desired.
\end{proof}

\subsection{Cliques are subspace-evasive}
Now, all that remains in the proof of our barrier result is to show that cliques are subspace-evasive (\cref{lem:main-subspace-evasive}). We do so below.

\begin{proof}[Proof of \cref{lem:main-subspace-evasive}]
We start with the standard observation that it suffices to prove a slightly stronger version of the result just for vector subspaces, since every affine subspace \(T\) of dimension \(\kappa\) is contained in a vector subspace \(S\) of dimension \(k\leq \kappa+1\) \cite{PR04}. Thus \begin{equation*}
    \frac{|\mathcal{Q} \cap T|}{|T|} > 2^{-(\kappa-3)/2} \implies \frac{|\mathcal{Q} \cap S|}{|S|} > 2^{-k/2+1}
\end{equation*}
so it suffices to show \(|\mathcal{Q}\cap S|/|S|\leq2^{-k/2+1}\) for every vector subspace \(S\) of dimension \(k\). Indeed, we proceed by proving this result.

Define $\mathcal{Q}_S:=\mathcal{Q}\cap S$. Notice that for any $x, y \in \mathcal{Q}_S$, we know that $x+y \in S$ as they each belong in $S$. By this observation, we see that $\mathcal{Q}_S + \mathcal{Q}_S \subseteq S$. Furthermore, we claim for any nonzero $z \in \mathcal{Q}_S + \mathcal{Q}_S$, there are at most $2$ pairs $(x,y) \in \mathcal{Q}_S \times \mathcal{Q}_S$ such that $z = x+y$. In other words, $\mathcal{Q}_S$ is \emph{a Sidon set}. Indeed, once we prove this, we find that $|\mathcal{Q}_S|(|\mathcal{Q}_S| - 1)/2 \le |\mathcal{Q}_S + \mathcal{Q}_S| - 1 \le |S|-1$. Thus $|\mathcal{Q}_S|^2/4 < |S|$, and so $|\mathcal{Q}_S|/|S|<2/\sqrt{|S|} = 2^{-k/2+1}$, which is what we wanted to show.

Consider the linear map $\varphi : \F_2^m \to \F_2^{k \times k}$ where the matrix $M = \varphi(Y)$ is defined by $M_{ij} = M_{ji} = Y_{ij}$. The key observation with this map is noticing that the set $\varphi(Q)$ corresponds exactly to the rank-1 matrices in $\text{im}(\varphi)$ as $\varphi(Y) = xx^\top$ for any $Y \in Q$. Now, for any nonzero vector $v \in Q_S + Q_S$, let us count the number of $y,y' \in \mathcal{Q}_S$ such that $v = y+y'$. By linearity and injectivity of $\varphi$, this is equivalent to counting the number of $x,x' \in \F_2^l$ satisfying $M_v = xx^\top + x'x'^\top$  where $M_v = \varphi(v)$.

Consider one such solution $M_v = zz^\top + z'z'^\top$. From the equation $M_v = zz^\top + z'z'^\top$, we see that the row space of $M_v$, which we denote by $\mathcal{R}(M_v)$, is contained in the span of $z$ and $z'$. Furthermore, since $M_v$ is nonzero, then $z \neq z'$. Thus at least two of the elements in $\{z,z', z+z'\}$ must be rows in $M_v$. This implies that the span of $z$ and $z'$ is contained in $\mathcal{R}(M_v)$. Hence $\mathcal{R}(M_v) = \{0, z, z', z+z'\}$. 

For any other such solution $M_v = xx^\top + x'x'^\top$, we also find by repeating the same steps that $x \neq x'$ and $\mathcal{R}(M_v) = \{0, x, x', x+x'\} = \{0, z, z', z+z'\}$. Let us now consider 2 possible cases:
\begin{enumerate}
    \item If either $z$ or $z'$ is zero, then without loss of generality, say $z'=0$. Then either $(x,x') = (z,0)$ or $(x,x') = (0,z)$. Thus there are at most $2$ possible pairs $(x,x')$ for this case.
    
    \item If both $z$ and $z'$ are nonzero, then both $x$ and $x'$ must be nonzero. Furthermore, $x, x'$ must be one of the elements in $\{z,z',z+z'\}$. From all $6$ possible pairs, only the pairs $(z,z')$ and $(z',z)$ satisfy the equation $M_v = xx^\top + x'x'^\top$. Thus there are at most $2$ possible pairs $(x,x')$ for this case.
\end{enumerate}
Since these are the only two possible cases, this therefore shows that $\mathcal{Q}_S$ is a Sidon set, which is what we wanted to show.
\end{proof}

\dobib

\bibliographystyle{alpha}
\bibliography{references}

\end{document}